\documentclass[11pt,letterpaper]{article}
\usepackage[T1]{fontenc}
\usepackage[margin=1in]{geometry}
\usepackage{amsmath}
\usepackage{newtxtext}
\usepackage[subscriptcorrection]{newtxmath}
\usepackage{graphicx}
\usepackage{xcolor}
\newcommand{\chg}[1]{{\color{black}#1}}
\usepackage{booktabs,longtable}
\newcommand{\tablefont}{\small}
\usepackage[round,authoryear]{natbib}
\usepackage[amsmath,thmmarks,hyperref]{ntheorem}
\theoremstyle{plain}
\theoremnumbering{arabic}
\theoremheaderfont{\normalfont\bfseries}
\theorembodyfont{\itshape}
\theoremseparator{.}
\theoremsymbol{}
\newtheorem{theorem}{Theorem}
\newtheorem{lemma}{Lemma}

\newtheorem{proposition}{Proposition}
\theorembodyfont{\normalfont}

\newtheorem{assumption}{Assumption}
\newtheorem{remark}{Remark}

\theoremstyle{nonumberplain}
\theoremheaderfont{\normalfont\itshape}
\theoremsymbol{\ensuremath{\square}}
\newtheorem{proof}{Proof}
\makeatletter
\renewtheoremstyle{nonumberplain}
  {\item[\theorem@headerfont\hskip\labelsep ##1\theorem@separator]}
  {\item[\theorem@headerfont\hskip\labelsep ##1\ ##3\theorem@separator]}
\makeatother
\usepackage[hidelinks]{hyperref}
\usepackage[capitalise,noabbrev]{cleveref}
\crefname{assumption}{Assumption}{Assumptions}
\hypersetup{pdftitle={Discovery-Targeted False Discovery Rate Control},pdfauthor={Zijun Gao}}
\newcommand{\FDP}{\operatorname{FDP}}
\newcommand{\E}{\mathrm{E}}
\newcommand{\var}{\operatorname{var}}
\newcommand{\ind}{\mathbf{1}}
\newif\ifprintproofs
\newenvironment{keywords}
  {\par\smallskip\noindent\small\textbf{Keywords: }}
  {\par\normalsize\medskip}
\printproofsfalse
\begin{document}
\title{Discovery-Targeted False Discovery Rate Control}
\author{Zijun Gao\\[.3em]\small Department of Data Sciences and Operations\\\small Marshall School of Business, University of Southern California}
\date{}
\maketitle
\begin{abstract}
In exploratory multiple testing, \chg{the} false discovery rate (FDR) is a commonly used error criterion, and existing methods typically \chg{take} an FDR level and return a set of discoveries. However, scientists often begin with a target number of discoveries, and accept any FDR level below a reasonable upper limit. Motivated by this practical need, we develop an FDR control approach with a data-driven, discovery-targeted level, which uses synthetic null p-values to split the p-value information between level selection and multiple testing. For independent p-values, the proposal controls finite-sample FDR given the reported level, which can therefore be interpreted in the standard way of a prespecified level. Asymptotically, the number of discoveries and the reported FDR level converge to the target and the optimal value. We also extend the approach to conformal p-values and include a null proportion estimate. In simulations and real-data examples, the procedure yields a number of discoveries close to the target and smaller FDR levels \chg{compared with} methods based on e-values or simultaneous false discovery proportion bounds.
\end{abstract}
\section{Introduction}
\label{sec:introduction}
High-throughput exploratory studies often test hundreds or thousands of hypotheses to identify candidates for follow-up validation experiments.
The false discovery rate (FDR) is a common measure of error in exploratory analysis, as it is less stringent than alternatives such as the familywise error rate (FWER).
Standard FDR control methods, such as the Benjamini--Hochberg (BH) procedure and its variants, start from a prespecified level and return a set of discoveries \citep{BenjaminiHochberg1995,StoreyTaylorSiegmund2004}.

\begin{figure}[!htb]
\centering
\includegraphics[width=.8\linewidth,trim=0bp 80bp 0bp 0bp,clip]{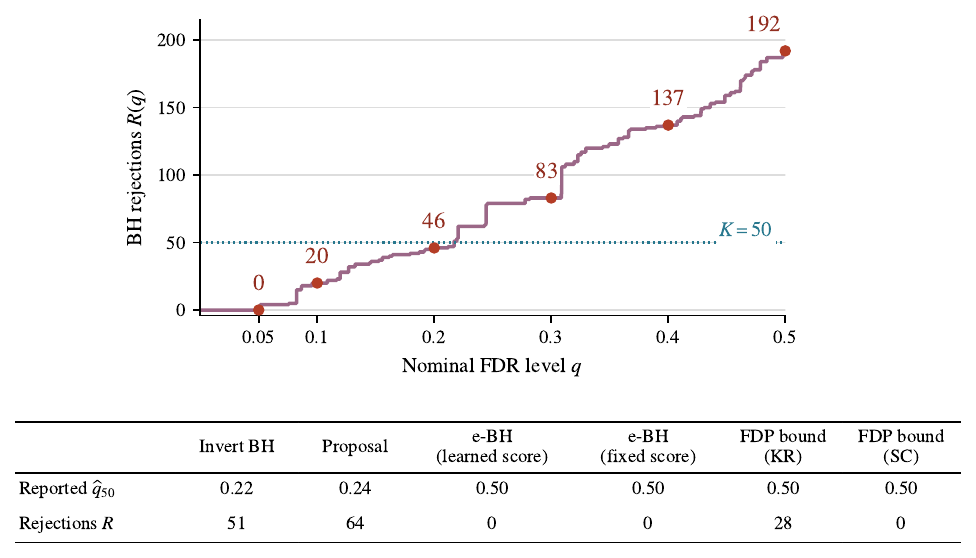}
\par\vspace{2pt}
\includegraphics[width=\linewidth,trim=0bp 0bp 0bp 190bp,clip]{generated/v5_figures/brca-student-bh-path-paper.pdf}
\caption{Discovery-targeted level selection for the breast-cancer data.
The plot shows the number of BH discoveries at FDR levels up to
$0.50$; the smallest BH level reaching the target $K=50$ is about
$0.22$.
The table gives the reported FDR level and the number of discoveries
for the methods of \cref{sec:numerical}, with $K=50$, $B=m$,
\chg{$[q_{\min},q_{\max}]=[0,0.50]$.}}
\label{fig:brca-bh-path}
\end{figure}

For scientists, the goal of exploratory analysis is to obtain a set of
candidates for downstream confirmatory studies, and the number of
discoveries can matter no less than the error rate: a certain number of discoveries are necessary for follow-up confirmation, whereas the FDR level
need only remain below an acceptable upper limit.
It has been recognized that a prespecified FDR level may not address the
demand \citep{Storey2002,GoemanSolari2011,KatsevichRamdas2020,%
XingEtAl2023,Grunwald2024,HemerikEtAl2024,XuEtAl2025,%
SongJinCandes2026,ZhuSimeone2026}.
As illustrated in \cref{fig:brca-bh-path} with the breast-cancer data
\citep{Hedenfalk2001,StoreyTibshirani2003}, for an investigator who
seeks around $50$ genes for follow-up and accepts FDR levels no larger than
$0.50$, the BH procedure at a
prespecified level of $0.05$ gives no discoveries, and hence no
candidates for a confirmatory study, whereas at $0.50$ it yields nearly
$200$, beyond the capacity for confirmation and at an unnecessarily
high error rate\footnote{Details of the data and tests are given in
\cref{sec:numerical}.}.

Motivated by this practical need, we aim for a procedure that allows
the investigator to specify a discovery target and returns a small
acceptable FDR level together with a rejection set of approximately
that size.
A tempting approach, which may have been adopted implicitly in practice,
raises the BH level until at least $K$ hypotheses are rejected and
adopts that level, or equivalently, uses the $K$th smallest
BH-adjusted p-value, or q-value, as the level
\citep{Storey2002,StoreyTibshirani2003}.
However, selecting the level from the same data that are then tested
need not control FDR (see Appendix~\ref{sec:bh-inversion-null} for a counter example).

In contrast to BH inversion, several lines of work provide valid error
control at data-driven levels that can target a given number of
discoveries.
\citet{XingEtAl2023} consider selecting a desired number of top-ranked
p-values, but the FDR guarantee holds only asymptotically.
Alternatively, simultaneous methods permit level selection by bounding
the FDP across rejection sets or thresholds
\citep{GoemanSolari2011,KatsevichRamdas2020,HemerikEtAl2024,%
SongJinCandes2026}; however, simultaneous control over all levels is unnecessary for a
single reported level and can sacrifice power.
In addition, e-value methods allow post hoc selection, including
data-dependent levels
\citep{WangRamdas2022,Grunwald2024,Koning2025,XuEtAl2025,ZhuSimeone2026};
however, the procedures are generally less powerful than their
p-value-based counterparts, owing to the conservativeness of Markov's inequality.

\begin{figure}[!htb]
\centering
\includegraphics[width=\linewidth]{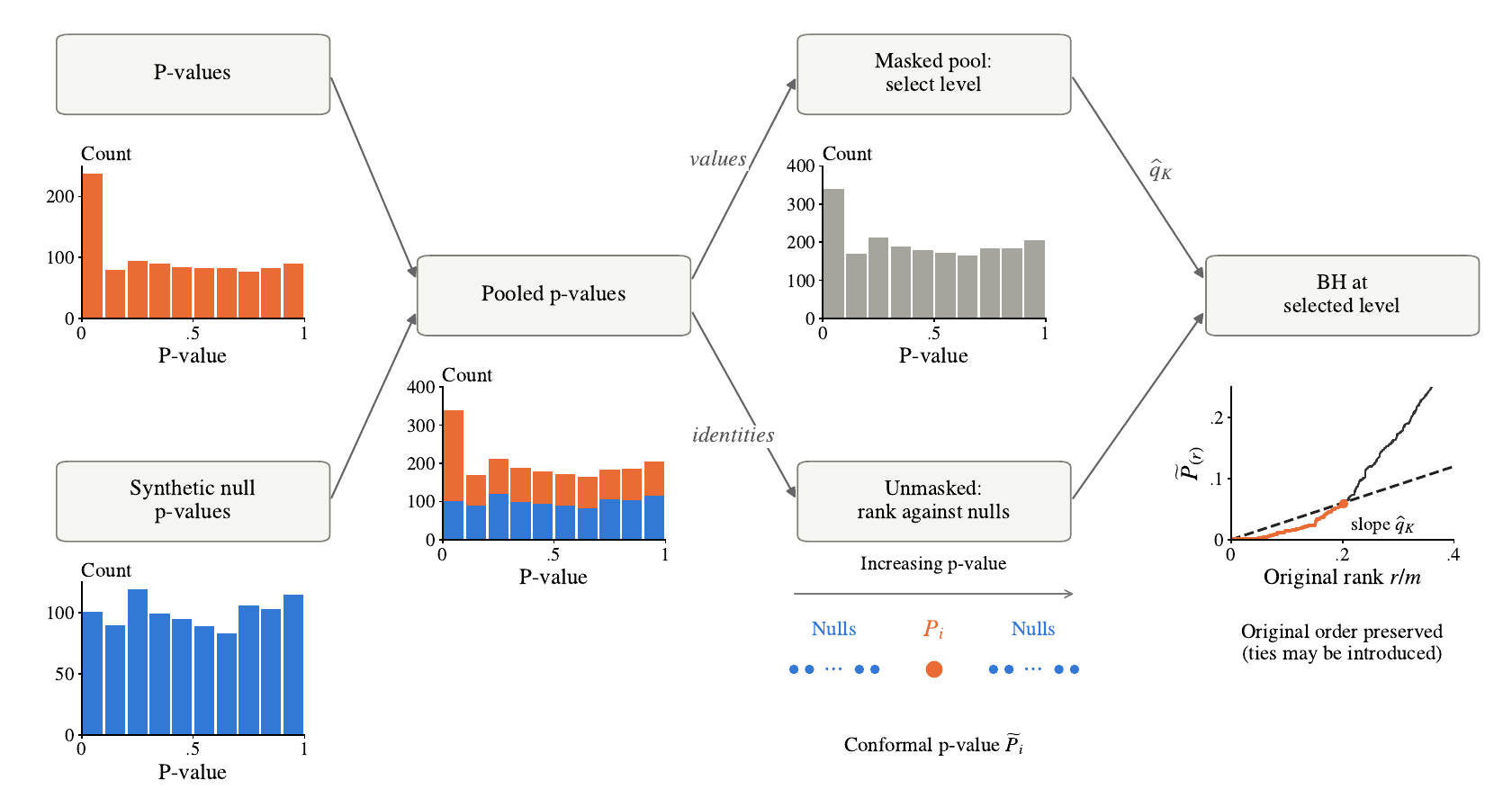}
\caption{Diagram of the proposal. The p-values are pooled with
synthetic null p-values, and the pool, with labels masked, determines
the level $\widehat q_K$. The labels are then unmasked to compute
conformal p-values, to which the BH procedure is applied at level
$\widehat q_K$.}
\label{fig:method-flow}
\end{figure}

We propose a procedure with finite-sample FDR control at a data-driven
level that targets the desired number of discoveries.
The proposal divides the information in the p-values between level
selection and multiple testing (\cref{fig:method-flow}).
Explicitly, following the null-augmentation idea in \citet{LiEtAl2026}, we
pool the p-values with synthetic null p-values and use the
identity-masked pool to select the FDR level.
We then unmask the identities to compute conformal p-values
\citep{BatesEtAl2023} and apply the BH procedure to them at the
selected level.
The procedure also extends to conformal
p-values \citep{BatesEtAl2023,GaoZhao2023} and can also incorporate an estimate of the null proportion
\citep{BenjaminiHochberg2000,Storey2002,StoreyTaylorSiegmund2004,%
BenjaminiKriegerYekutieli2006,Gao2025}.
\begin{itemize}
    \item For FDR control, let $S$ and $\widehat q_K$ denote the rejection set
and the selected level from the procedure, we show that
\begin{equation}
\E\{\FDP(S)\mid\widehat q_K\}\le\widehat q_K,
\label{eq:target-guarantee}
\end{equation}
where $\FDP(S)$ is the false discovery proportion of $S$.
The conditional guarantee implies that,
although data-dependent, $\widehat q_K$ admits the same interpretation
as a prespecified level; \eqref{eq:target-guarantee} also implies the
marginal guarantee $\E\{\FDP(S)/\widehat q_K\}\le 1$ of e-value methods
\citep{XuEtAl2025,ZhuSimeone2026}, but not conversely.
\item 
For the number of rejections and the selected level, we show asymptotic
optimality, in the sense that the number of rejections concentrates
around $K$ and $\widehat q_K$ converges to its oracle value (the smallest level that yields $K$ rejections in the population).
\end{itemize}
\chg{In the motivating example, the proposal makes 64 discoveries at reported level 0.24, whereas the FDP-bound method of \citet{KatsevichRamdas2020} makes 28 and the other FDP-bound method and the e-value methods make none, even at the cap of 0.50 (details in
\cref{sec:numerical}).}

\noindent\textbf{Organization.}
\Cref{sec:formulation} formulates the problem.
\Cref{sec:method} describes the proposal.
\Cref{sec:bag-choice} establishes conditional FDR control and
asymptotic optimality.
\Cref{sec:numerical} contains simulations and the breast-cancer
example.
\chg{The Appendix} gives proposal extensions, proofs, and additional empirical results.

\noindent\textbf{Notation.}
For a vector $v$, $v_{(r)}$ is its $r$th smallest coordinate.
The symbol $\ind\{A\}$ denotes the indicator of an event $A$.
We write $a\vee b=\max(a,b)$ and $x_+=\max(x,0)$, and use
$[x]_{[a,b]}=\min\{b,\max(a,x)\}$ for clipping.
Let $\Phi$ denote the standard normal distribution function.
For a square-integrable random variable $Y$,
$\|Y\|_2=\{\E(Y^2)\}^{1/2}$.
The relation $a_m\asymp b_m$ means that their positive ratio is bounded above and away from zero.

\section{Problem formulation}
\label{sec:formulation}
Let $P_1,\ldots,P_m$ be independent p-values, let $\mathcal H_0$ be the
fixed set of true null hypotheses, and write $m_0=|\mathcal H_0|$, $\pi_0=|\mathcal H_0|/m$.
Assume $P_i$ is uniform on $(0,1)$ for $i\in\mathcal H_0$
(\chg{\cref{rem:conservative-nulls}} extends to null p-values that are stochastically larger than uniform).
For a rejection set $S$, define
\[
\operatorname{FDR}(S)=\E\{\FDP(S)\},\quad
\FDP(S)=\frac{|S\cap\mathcal H_0|}{R\vee1},\quad R=|S|.
\]
Let $1\le K<m$ be a prespecified target number of discoveries and
$[q_{\min},q_{\max}]$ a prespecified range of acceptable levels, where
$0\le q_{\min}<q_{\max}<1$ ($q_{\min}=0$ imposes no lower limit).
Our goal is to select a level $\widehat q_K\in[q_{\min},q_{\max}]$ and
a rejection set $S$ such that $R$ is close to $K$ and the conditional
guarantee \eqref{eq:target-guarantee} holds.

For p-values $p=(p_1,\ldots,p_m)$ with ordered values
$p_{(1)}\le\cdots\le p_{(m)}$, the BH procedure at level $q\in[0,1)$
rejects the set $S_q(p)$ of size $R_q(p)$, where
\[
R_q(p)=\max\{r\in\{1,\ldots,m\}:p_{(r)}\le qr/m\},
\quad
S_q(p)=\{i:p_i\le p_{(R_q(p))}\},
\]
with $R_q(p)=0$ and $S_q(p)=\emptyset$ if no such $r$ exists.

\section{Method}
\label{sec:method}

In this section, we describe our proposal.
We generate synthetic null p-values and pool them with the original p-values.
The pooled p-values contain two types of information: the set of their
values, and the labels of these values (whether the value is an
original p-value or a synthetic null).
We use the first to select the FDR level and the second for multiple testing.
Unlike randomly splitting the hypotheses, the procedure keeps every hypothesis in the
testing step. Below we provide the details.

For a fixed integer $B\ge1$, we generate synthetic null p-values
$U_1,\ldots,U_B$, independent and uniform on $(0,1)$, and pool all
$m+B$ values with their labels masked.
Let $\widehat F_{\mathrm{all}}$ be the empirical distribution function
of the pool.
Of the $(m+B)\widehat F_{\mathrm{all}}(t)$ pooled values not exceeding
$t$, about $Bt$ are synthetic, so
$(m+B)\widehat F_{\mathrm{all}}(t)-Bt$ estimates the number of original
p-values not exceeding $t$ without using the labels.
We take the smallest threshold at which this estimate reaches $K$,
\begin{equation}
\widehat t_{K,B}
=\inf\{t\in[0,1]:(m+B)\widehat F_{\mathrm{all}}(t)-Bt\ge K\}.
\label{eq:pooled-threshold}
\end{equation}
We then define the level by analogy with the q-value at rank $K$, with
$\widehat t_{K,B}$ in place of the $K$th smallest p-value:
\begin{equation}
\widehat q^{\mathrm{raw}}_{K,B}
=\frac mK\,\frac{1+B\widehat t_{K,B}}{B+1},
\quad
\widehat q_K=\chg{[\widehat q^{\mathrm{raw}}_{K,B}]_{[q_{\min},q_{\max}]}}.
\label{eq:pooled-level}
\end{equation}
Unlike BH inversion, $\widehat q_K$ depends on the data only through
the masked pool.

\begin{remark}
\label{rem:other-estimates}
The estimate $(m+B)\widehat F_{\mathrm{all}}(t)-Bt$ in \eqref{eq:pooled-threshold} may be replaced by any other estimate of the number of original p-values not exceeding $t$, provided that it depends on the data only through the masked pool.
\end{remark}

Once $\widehat q_K$ is obtained, we unmask the labels and use the $B$
synthetic nulls to compute the conformal p-values and the rejection set
\begin{equation}
\widetilde P_i=\frac{1+\sum_{b=1}^B\ind\{U_b\le P_i\}}{B+1},
\quad
S=S_{\widehat q_K}(\widetilde P),
\label{eq:selection}
\end{equation}
where $\widetilde P=(\widetilde P_1,\ldots,\widetilde P_m)$; that is,
the BH procedure is applied to the conformal p-values at level
$\widehat q_K$.
Since $\widetilde P_i$ is a nondecreasing function of $P_i$, the
conformal p-values preserve the ordering of the original p-values, and
$S$ always consists of the hypotheses with the smallest original
p-values.

The number $B$ of synthetic nulls balances two sources of error: a
larger $B$ yields finer conformal p-values but a more variable estimated level
$\widehat q_K$.
\chg{For the asymptotic analysis, we consider the marginal model}
\begin{equation}
P_1,\ldots,P_m\stackrel{\text{i.i.d.}}{\sim} F,
\label{eq:model}
\end{equation}
where $F$ is a distribution function with density $f$.
Let $t_K$ be the smallest solution of $F(t)=K/m$, and suppose that
$mt_Kf(t_K)/K<1$.
In \chg{\cref{prop:variant-moments}}, we show that
\begin{equation}
\E(R)\approx K,\quad
\var(R)\approx
\frac{t_K(1-t_K)}{\{1-mt_Kf(t_K)/K\}^2}
\left\{B+2B^\star+\frac{(B^\star)^2}{B}\right\},
\label{eq:count-variance}
\end{equation}
where $B^\star=mf(t_K)$.
\chg{The variance is minimized at $B=B^\star$.}
Since $B^\star$ depends on the unknown density $f(t_K)$, we use $B=m$
as a default, which is likely to be close to the optimal
choice when the null proportion is close to one.

We extend the procedure to incorporate a null proportion estimate in
Appendix~\ref{sec:adaptive-extension} and to conformal p-values in
Appendix~\ref{sec:conformal-extension}, and \cref{rem:conservative-nulls} extends its guarantee to null p-values that are stochastically larger than uniform.

\section{FDR control and asymptotic optimality}
\label{sec:bag-choice}

\subsection{Conditional FDR control}

\begin{theorem}[Conditional FDR control]
\label{thm:main-control}
Suppose that $P_1,\ldots,P_m$ are independent and that $P_i$ is uniform
on $(0,1)$ for $i\in\mathcal H_0$.
Then, for any $B\ge1$, the rejection set $S$ and the level
$\widehat q_K$ defined by \eqref{eq:pooled-level} and
\eqref{eq:selection} satisfy
\[
\E\{\FDP(S)\mid\widehat q_K\}\le\frac{m_0}{m}\,\widehat q_K
\quad\text{a.s}.
\]
\end{theorem}

The conditional guarantee means that $\widehat q_K$ admits the same
interpretation as a prespecified level; it also implies the marginal
guarantee $\E\{\FDP(S)/\widehat q_K\}\le1$.
Similar results for the adaptive extension and for conformal p-values
are given in Appendices~\ref{sec:adaptive-extension}
and~\ref{sec:conformal-extension}.

\chg{\begin{remark}
\label{rem:conservative-nulls}
\Cref{thm:main-control} also holds if each null p-value $P_i$, $i\in\mathcal H_0$, has a nondecreasing density $f_i$ on $(0,1)$ instead of being uniform.
\end{remark}}

\subsection{Asymptotic optimality}

We show the asymptotic optimality of the procedure when $B$ is of the
same order as $m$, including the default choice $B=m$.
Under model \eqref{eq:model}, let $t_K$ be the solution of
$F(t)=K/m$.
Let $q_*=mt_K/K$, which we call the oracle level, as it is the smallest
level at which the BH procedure rejects at least $K$ hypotheses in the
population when $F$ is concave.

\begin{assumption}
\label{ass:regular}
As $m\to\infty$, $K/m\to\kappa\in(0,1)$ and $B/m\to\beta\in(0,\infty)$.
Let $t_\kappa$ be the solution of $F(t)=\kappa$.
The density $f$ is continuous at $t_\kappa$, and
$0<\tfrac{t_\kappa f(t_\kappa)}{\kappa}<1$, \chg{$q_{\min}<t_\kappa/\kappa<q_{\max}$}.
\end{assumption}

\begin{proposition}[Asymptotic optimality]
\label{prop:targeting}
Under \Cref{ass:regular}, as $m\to\infty$,
\[
m^{1/2}(1-R/K)_+=O_p(1),\quad
m^{1/2}\,\bigl\|\widehat q_K-q_*\bigr\|_2\to
\frac{\{\kappa(1-\kappa)+\beta t_\kappa(1-t_\kappa)\}^{1/2}}
{\kappa f(t_\kappa)}.
\]
If, in addition, $F$ is concave, then
\[
m^{1/2}\left\|\frac RK-1\right\|_2\to
\frac{\{\beta+f(t_\kappa)\}\{t_\kappa(1-t_\kappa)\}^{1/2}}
{\beta^{1/2}\{\kappa-t_\kappa f(t_\kappa)\}}.
\]
\end{proposition}

\section{Numerical results}
\label{sec:numerical}
\subsection{Simulation}
\begin{figure}[tbp]
\centering
\includegraphics[width=\textwidth]{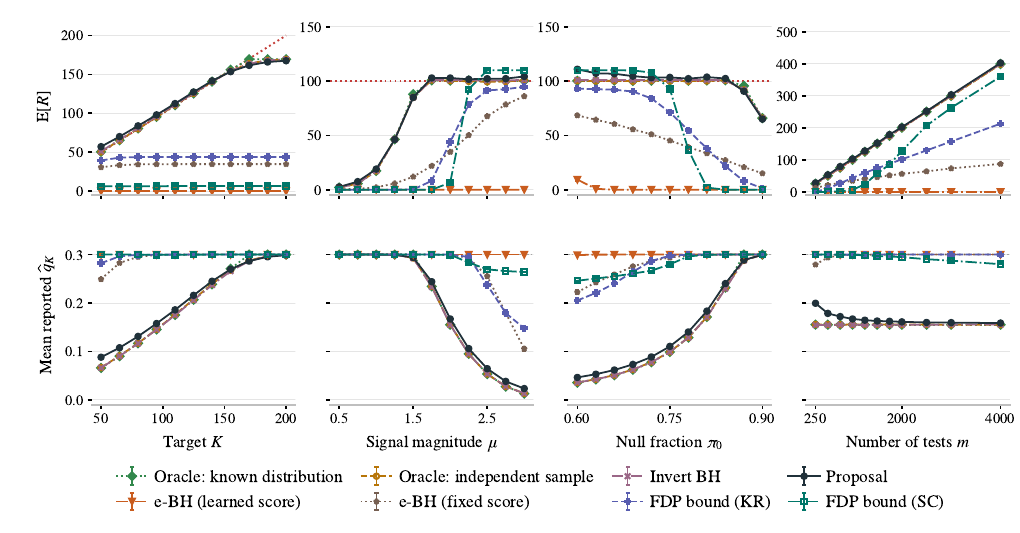}
\caption{\chg{Mean number of rejections (top) and mean reported level (bottom) as $K$, $\mu$, $\pi_0$ and $m$ vary, with the other parameters at the default $(m,B,K,\mu,\pi_0)=(1000,1000,100,2,0.8)$; the size sweep uses $B=m$ and $K=m/10$.
Levels are clipped to $[0.01,0.30]$.
Curves are means over 3000 paired repetitions, including empty outputs; bars show $1.96$ Monte Carlo standard errors.
Details of the methods are given in Appendix~\ref{sec:simulation-details}.}}
\label{fig:simulation-panels}
\end{figure}

We compare the following methods (more details in
Appendix~\ref{sec:simulation-details}).
\begin{itemize}
\item \textit{Proposal:} procedure in \cref{sec:method} with the recommended
$B=m$.
\item \textit{Oracles:} the BH procedure at the oracle level $q_*$; and the BH procedure at a level
estimated from an additional independent sample of $m$ hypotheses.
\item \textit{BH inversion \chg{(Invert BH in the plots)}:} the BH procedure at the \chg{$K$th} smallest
BH-adjusted p-value (no FDR control).
\item \textit{E-value methods:} conformal e-BH
\citep{ZhuSimeone2026} with either a \chg{learned or a fixed score}.
\item \textit{FDP-bound methods:} simultaneous bounds of
\citet{KatsevichRamdas2020} and of \citet{SongJinCandes2026} at confidence level
99\% converted to provide
\chg{the marginal guarantee $\E\{\FDP(S)/\widehat q_K\}\le1$}.
\end{itemize}

We generate independent scores $X_i\sim N(0,1)$ for
$i\in\mathcal H_0$ and $X_i\sim N(-\mu,1)$ otherwise, and set
$P_i=\Phi(X_i)$.
The default is $(m,B,K,\mu,\pi_0)=(1000,1000,100,2,0.8)$.
\chg{Reported levels are restricted to $[q_{\min},q_{\max}]=[0.01,0.30]$.
\Cref{fig:simulation-panels} varies $K$, $\mu$, $\pi_0$ and $m$, with $B=m$ and $K=m/10$.
Each setting uses 3000 repetitions.}

\Cref{fig:simulation-panels} reports the number of rejections and the
reported level.
At the default setting, the proposal and the independent-sample oracle
make 102.9 and 100.4 discoveries on average, at mean levels 0.167 and
0.156, respectively.
The proposal largely follows the oracle curves, although the cap at
$q_{\max}$ limits the number of discoveries for large targets
($K>170$), weak signals ($\mu<1.6$) or high null proportions
($\pi_0>0.86$).
The e-value and FDP-bound methods often report higher FDR levels, near
the cap, while still making fewer discoveries.

\subsection{Breast-cancer dataset}
We analyse the hereditary breast-cancer data of \citet{Hedenfalk2001},
as reanalysed by \citet{StoreyTibshirani2003}, which give
log-base-two expression for 3170 genes in seven BRCA1 and eight BRCA2
tumours.
For each gene we compute a one-sided two-sample Student t-test for
lower expression in BRCA1 tumours, with 13 degrees of freedom.
\chg{We set $K=50$, $B=m=3170$ and $[q_{\min},q_{\max}]=[0.01,0.50]$.}
The results are given in the table of \cref{fig:brca-bh-path}.
The proposal makes 64 discoveries at reported level 0.2386.
\chg{Both e-value methods and the SC bound make no discoveries at $q_{\max}$, whereas the KR bound makes 28.}

\section{Conclusion}
\label{sec:discussion}

In exploratory studies, investigators often need a certain number of discoveries as well as FDR control.
In this paper, we address this need with a procedure that aims at a target number of discoveries and reports the corresponding FDR level.
We use synthetic null p-values to split the information between label-blind level selection and label-dependent testing, which yields FDR control conditionally on the reported level.
Asymptotically, the number of rejections and the reported level converge to the target and the optimal value.

\vspace{0.5cm}
\par\smallskip\noindent\chg{\textit{AI disclaimer.}
OpenAI ChatGPT and Anthropic Claude were used to assist with coding, proofs and drafting.
The author takes full responsibility for the content of the paper.}

\begingroup
\setlength{\bibsep}{2pt}
\bibliography{references}
\endgroup

\appendix


\section*{Appendices}
\phantomsection
\addcontentsline{toc}{section}{Appendices}
\chg{These appendices give the extensions of the procedure and their guarantees (\cref{sec:extensions}), the proofs (\cref{sec:proofs}) and additional numerical results (\cref{sec:simulation-details}).}
We use the main article's notation: $m$ hypotheses, true-null set $\mathcal H_0$ of size $m_0$, target $K$, calibration size $B$, and selected set $S$ of size $R$.
Thus $\FDP(S)=|S\cap\mathcal H_0|/(R\vee1)$; $\Phi$ denotes the standard normal distribution function and $\|X\|_p=(\E|X|^p)^{1/p}$.



\section{Extensions}
\label{sec:extensions}

\subsection{Extension to a null proportion estimate}
\label{sec:adaptive-extension}

\subsubsection{Method}
\label{sec:sup-algorithm}

Fix $\lambda\in(0,1)$.
Of the $(m+B)\{1-\widehat F_{\mathrm{all}}(\lambda)\}$ pooled values
exceeding $\lambda$, about $B(1-\lambda)$ are synthetic, so
$(m+B)\{1-\widehat F_{\mathrm{all}}(\lambda)\}-B(1-\lambda)$ estimates
the number of original p-values exceeding $\lambda$ without using the
labels.
This gives a masked-pool version of the estimator of
\citet{Storey2002},
\begin{equation}
\widehat\pi_0^{\mathrm{pool}}
=\min\left[1,\max\left\{0,
\frac{1+(m+B)\{1-\widehat F_{\mathrm{all}}(\lambda)\}-B(1-\lambda)}
{m(1-\lambda)}\right\}\right].
\label{eq:pooled-pi0}
\end{equation}
We then define the FDR level by multiplying the level in
\eqref{eq:pooled-level} by this null proportion estimate:
\begin{equation}
\widehat q^{\mathrm{ada}}_K
=\chg{\bigl[\widehat\pi_0^{\mathrm{pool}}\,
\widehat q^{\mathrm{raw}}_{K,B}\bigr]_{[q_{\min},q_{\max}]}}.
\label{eq:adaptive-level}
\end{equation}
Like $\widehat q_K$, the level $\widehat q^{\mathrm{ada}}_K$ depends on the data only through the masked pool.

Once $\widehat q^{\mathrm{ada}}_K$ is obtained, we unmask the labels and
compute the conformal p-values in \eqref{eq:selection}, with ordered
values $\widetilde P_{(1)}\le\cdots\le\widetilde P_{(m)}$.
For a fixed integer $1\le k_\tau\le B$ and $\tau=k_\tau/(B+1)$, we
estimate the null proportion again and apply the adaptive BH procedure
of \citet{BatesEtAl2023}:
\begin{equation}
\begin{aligned}
R^{\mathrm{ada}}
=\max\left\{r:\widetilde P_{(r)}\le
\frac{\widehat q^{\mathrm{ada}}_Kr}{m\widehat\pi_0^{\mathrm{conf}}},
\ \widetilde P_{(r)}<\tau\right\},~~\widehat\pi_0^{\mathrm{conf}}
=\frac{1+\sum_{i=1}^m\ind\{\widetilde P_i\ge\tau\}}{m(1-\tau)},~~
S^{\mathrm{ada}}=\{i:\widetilde P_i\le\widetilde P_{(R^{\mathrm{ada}})}\},
\end{aligned}
\label{eq:adaptive-rule}
\end{equation}
with $R^{\mathrm{ada}}=0$ and $S^{\mathrm{ada}}=\emptyset$ if no such
$r$ exists. We remark that the estimate $\widehat\pi_0^{\mathrm{conf}}$ is not
truncated at one.

The number $B$ of synthetic nulls again balances finer conformal
p-values against a more variable estimated level.
Under model \eqref{eq:model}, suppose that $t_K<\lambda$, and let
\[
\rho=\frac{1-F(\lambda)}{1-\lambda},\quad
w=\frac{t_Kf(t_K)}{\rho(1-\lambda)},
\]
where $\rho$ is the limit of the estimate of the null proportion.
In \cref{prop:variant-moments}, we show that, as $m\to\infty$ with $K$
and $B$ of the same order as $m$,
\begin{equation}
\E(R^{\mathrm{ada}})=K+o(m^{1/2}),\quad
\var(R^{\mathrm{ada}})=
\frac{c_0B+2mc_1+m^2c_2/B}{\{1-mt_Kf(t_K)/K\}^2}+o(m),
\label{eq:adaptive-count-variance}
\end{equation}
where
\begin{equation}
\begin{aligned}
c_0&=t_K(1-t_K)+2wt_K(1-\lambda)+w^2\lambda(1-\lambda),\\
c_1&=f(t_K)t_K(1-t_K)+w\{f(t_K)+f(\lambda)\}t_K(1-\lambda)
+w^2f(\lambda)\lambda(1-\lambda),\\
c_2&=f(t_K)^2t_K(1-t_K)+2wf(t_K)f(\lambda)t_K(1-\lambda)
+w^2f(\lambda)^2\lambda(1-\lambda).
\end{aligned}
\label{eq:adaptive-coefficients}
\end{equation}
The leading term of the variance is minimized at
$B=m(c_2/c_0)^{1/2}$.
Since this value depends on unknown quantities \chg{such as $t_K$ and $f(t_K)$}, we again use $B=m$ as a practical default.
We label the resulting procedure \textit{Proposal + adaptive BH} in the numerical comparisons.

\subsubsection{Conditional FDR control}

\begin{theorem}[Conditional FDR control]
\label{thm:adaptive-control}
Suppose that $P_1,\ldots,P_m$ are independent and that $P_i$ is uniform
on $(0,1)$ for $i\in\mathcal H_0$.
Then, for any $B\ge1$, $\lambda\in(0,1)$ and $1\le k_\tau\le B$, the
rejection set $S^{\mathrm{ada}}$ and the level
$\widehat q^{\mathrm{ada}}_K$ defined by \eqref{eq:adaptive-level} and
\eqref{eq:adaptive-rule} satisfy
\[
\E\{\FDP(S^{\mathrm{ada}})\mid\widehat q^{\mathrm{ada}}_K\}
\le\widehat q^{\mathrm{ada}}_K
\quad\text{a.s}.
\]
\end{theorem}

Compared with \cref{thm:main-control}, the bound no longer carries the
factor $m_0/m$ \chg{because the null proportion is estimated}.
The proof is given in \cref{sec:proofs}.

{\printproofsfalse
\long\gdef\ProofOrdinaryControl{%
\begin{theorem}[Conditional FDR control for a general level]
\label{thm:conditional-control}
Let $\mathcal G$ be the sigma-field generated by the order statistics
of $(P_1,\ldots,P_m,U_1,\ldots,U_B)$.
Under the assumptions of \cref{thm:main-control}, for any
$\mathcal G$-measurable $\widehat q\in[0,1)$ and
$S=S_{\widehat q}(\widetilde P)$,
\[
\E\{\FDP(S)\mid\widehat q\}\le\frac{m_0}{m}\,\widehat q
\quad\text{a.s}.
\]
\end{theorem}

\begin{proof}[of \cref{thm:conditional-control}]
Fix $i\in\mathcal H_0$.
Let $W_{(1)}<\cdots<W_{(B+1)}$ be the order statistics of
$P_i,U_1,\ldots,U_B$, which are distinct almost surely, and let $\rho_i$
be the rank of $P_i$ among them.
Let $\mathcal F_i$ be the sigma-field generated by $(P_j)_{j\ne i}$ and
$(W_{(1)},\ldots,W_{(B+1)})$.
Since $P_i,U_1,\ldots,U_B$ are independent and uniform on $(0,1)$, and
independent of $(P_j)_{j\ne i}$, the rank $\rho_i$ is uniform on
$\{1,\ldots,B+1\}$ and independent of $\mathcal F_i$.
The masked pool is a function of $(P_j)_{j\ne i}$ and
$(W_{(1)},\ldots,W_{(B+1)})$, so $\mathcal G\subseteq\mathcal F_i$ and
$\widehat q$ is $\mathcal F_i$-measurable.

Given $\mathcal F_i$, write $\widetilde P_j(r)$ and $R(r)$ for the
conformal p-values and the number of rejections when $\rho_i=r$.
Then $\widetilde P_i(r)=r/(B+1)$ and, for $j\ne i$,
\[
\sum_{b=1}^B\ind\{U_b\le P_j\}
=\sum_{s=1}^{B+1}\ind\{W_{(s)}\le P_j\}-\ind\{W_{(r)}\le P_j\},
\]
which is nondecreasing in $r$.
Hence every conformal p-value is nondecreasing in $r$, and $R(r)$ is
nonincreasing in $r$.
By the definition of the BH procedure, hypothesis $i$ is rejected when
$\rho_i=r$ if and only if
\begin{equation}
\frac r{B+1}\le\frac{\widehat q\,R(r)}m .
\label{eq:proof-reject}
\end{equation}
The left side of \eqref{eq:proof-reject} is increasing in $r$ and the
right side is nonincreasing, so \eqref{eq:proof-reject} holds exactly
for $r\le r_*$, for some $r_*\in\{0,\ldots,B+1\}$.
If $r_*\ge1$,
\[
\E\left\{\left.\frac{\ind\{i\in S\}}{R\vee1}\right|\mathcal F_i\right\}
=\frac1{B+1}\sum_{r=1}^{r_*}\frac1{R(r)}
\le\frac{r_*}{(B+1)R(r_*)}\le\frac{\widehat q}{m},
\]
where the equality holds because $\rho_i$ is uniform and independent of
$\mathcal F_i$, the first inequality because $R(r)\ge R(r_*)$ for
$r\le r_*$, and the second by \eqref{eq:proof-reject} at $r=r_*$.
If $r_*=0$ the left side is zero.
Since $\mathcal G\subseteq\mathcal F_i$, the same bound holds given
$\mathcal G$, and summing over $i\in\mathcal H_0$ gives
$\E\{\FDP(S)\mid\mathcal G\}\le(m_0/m)\widehat q$.
\end{proof}
}
\ifprintproofs\ProofOrdinaryControl\fi

\long\gdef\ProofAdaptiveControl{%
\begin{proof}[of \cref{thm:adaptive-control}]
Let $\mathcal G$ be as in \cref{thm:conditional-control}.
As $\widehat q^{\mathrm{ada}}_K$ is $\mathcal G$-measurable, it suffices
to show that
$\E\{\FDP(S^{\mathrm{ada}})\mid\mathcal G\}\le\widehat q$ for any
$\mathcal G$-measurable $\widehat q\in[0,1)$ in place of
$\widehat q^{\mathrm{ada}}_K$ in \eqref{eq:adaptive-rule}.
If $m_0=0$ this is immediate, and if $\widehat q=0$ or $k_\tau=1$ no
hypothesis is rejected; we therefore take $m_0\ge1$, $\widehat q>0$ and
$k_\tau\ge2$.

Fix $i\in\mathcal H_0$, with $W_{(1)}<\cdots<W_{(B+1)}$, $\rho_i$ and
$\mathcal F_i$ as in the proof of \cref{thm:conditional-control}, and
let $\mathcal H$ be the sigma-field generated by
$(P_j)_{j\notin\mathcal H_0}$ and the order statistics of
$\{P_j:j\in\mathcal H_0\}\cup\{U_1,\ldots,U_B\}$, so that
$\mathcal G\subseteq\mathcal H\subseteq\mathcal F_i$.
Given $\mathcal F_i$, write $A(r)$ and $R(r)$ for
$\sum_{j=1}^m\ind\{\widetilde P_j\ge\tau\}$ and $R^{\mathrm{ada}}$ when
$\rho_i=r$.
Every conformal p-value is nondecreasing in $r$, so $A(r)$ is
nondecreasing and $R(r)$ is nonincreasing, and
$i\in S^{\mathrm{ada}}$ if and only if
\begin{equation}
\frac r{B+1}\le\frac{\widehat q\,(1-\tau)R(r)}{1+A(r)},
\quad
\frac r{B+1}<\tau ,
\label{eq:proof-reject-ada}
\end{equation}
that is, if and only if $r\le r_*$ for some $r_*\in\{0,\ldots,B+1\}$.
For $r_*\ge1$,
\begin{equation}
\E\left\{\left.
\frac{\ind\{i\in S^{\mathrm{ada}}\}}{R^{\mathrm{ada}}\vee1}
\right|\mathcal F_i\right\}
=\frac1{B+1}\sum_{r=1}^{r_*}\frac1{R(r)}
\le\frac{r_*}{(B+1)R(r_*)}
\le\frac{\widehat q\,(1-\tau)}{1+A(1)},
\label{eq:proof-ada-bound}
\end{equation}
by \eqref{eq:proof-reject-ada} at $r=r_*$ and $A(r_*)\ge A(1)$; for
$r_*=0$ the left side is zero.

Let $A=A(\rho_i)$, and let $V$ be the $(k_\tau-1)$th smallest of
$U_1,\ldots,U_B$.
Then $\widetilde P_j\ge\tau$ if and only if $P_j\ge V$, so
$A\ge Y=\sum_{j\in\mathcal H_0,\,j\ne i}\ind\{P_j\ge V\}$.
Since $\rho_i$ is uniform and independent of
$\mathcal F_i\supseteq\mathcal H$,
\begin{equation}
\begin{aligned}
\E\bigl[\{1+A(1)\}^{-1}\mid\mathcal H\bigr]
&=(B+1)\,\E\bigl[(1+A)^{-1}\ind\{\rho_i=1\}\mid\mathcal H\bigr]\\
&\le(B+1)\,\E\bigl[(1+Y)^{-1}\ind\{\rho_i=1\}\mid\mathcal H\bigr]\\
&=\E\bigl\{(1+Y)^{-1}\mid\rho_i=1\bigr\},
\end{aligned}
\label{eq:proof-ada-rank1}
\end{equation}
where the last equality holds because, given $\mathcal H$, every
assignment of the $m_0+B$ null and synthetic null values to their
labels is equally likely, and $Y$ and $\ind\{\rho_i=1\}$ depend only on
the relative order of these values.
Given $\rho_i=1$, $V=W_{(k_\tau)}$ has the
$\mathrm{Beta}(k_\tau,B+2-k_\tau)$ distribution and, given $V$, $Y$ is
binomial with index $m_0-1$ and probability $1-V$.
Hence
\begin{equation}
\begin{aligned}
\E\bigl\{(1+Y)^{-1}\mid\rho_i=1\bigr\}
&=\E\left\{\frac{1-V^{m_0}}{m_0(1-V)}\right\}
\le\frac1{m_0}\,\E\left(\frac1{1-V}\right)\\
&=\frac{B+1}{m_0(B+1-k_\tau)}
=\frac1{m_0(1-\tau)}.
\end{aligned}
\label{eq:proof-ada-beta}
\end{equation}
By \eqref{eq:proof-ada-bound}--\eqref{eq:proof-ada-beta},
$\E[\ind\{i\in S^{\mathrm{ada}}\}/(R^{\mathrm{ada}}\vee1)
\mid\mathcal H]\le\widehat q/m_0$.
Summing over $i\in\mathcal H_0$ and taking expectations given
$\mathcal G$ gives the result.
\end{proof}
}
\ifprintproofs\ProofAdaptiveControl\fi
}

\subsubsection{Asymptotic optimality}

Let $q^{\mathrm{ada}}_*=\rho\,mt_K/K$, which we call the oracle level of
the adaptive procedure, as it is the smallest level at which the BH
procedure with null proportion $\rho$ rejects at least $K$ hypotheses
in the population when $F$ is concave.

\begin{proposition}[Asymptotic optimality]
\label{prop:adaptive-targeting}
Suppose that \Cref{ass:regular} holds, and further assume that the
density $f$ is continuous at $\lambda$, $t_\kappa<\lambda$ and
$0<\rho<1$. \chg{Assume also that $\tau=\lambda+O(B^{-1})$ and $q_{\min}<\rho t_\kappa/\kappa<q_{\max}$.}
Let $w$, $c_0$, $c_1$ and $c_2$ be as in
\eqref{eq:adaptive-coefficients} with $t_K$ replaced by $t_\kappa$, and
let
$v=\kappa(1-\kappa)+2w\kappa\{1-F(\lambda)\}
+w^2F(\lambda)\{1-F(\lambda)\}$.
Then, as $m\to\infty$,
\[
m^{1/2}(1-R^{\mathrm{ada}}/K)_+=O_p(1),\quad
m^{1/2}\,\bigl\|\widehat q^{\mathrm{ada}}_K-q^{\mathrm{ada}}_*\bigr\|_2
\to\frac{\rho\,(v+\beta c_0)^{1/2}}{\kappa f(t_\kappa)}.
\]
If, in addition, $F$ is concave, then
\[
m^{1/2}\left\|\frac{R^{\mathrm{ada}}}K-1\right\|_2\to
\frac{(\beta c_0+2c_1+c_2/\beta)^{1/2}}
{\kappa-t_\kappa f(t_\kappa)}.
\]
\end{proposition}

Compared with \cref{prop:targeting}, the same target is met at an
oracle level smaller by the factor $\rho$.

\subsection{Extension to conformal p-values}
\label{sec:conformal-extension}

\subsubsection{Method}

\chg{Let $T_1,\ldots,T_m$ be test scores obtained from a fixed scoring rule, with small values providing}
evidence against the null, and suppose that their null distribution
function $F_0$ is unknown but a null reference sample is available
\citep{MaryRoquain2022,BatesEtAl2023,GaoZhao2023}.
We split the reference sample into an
estimation fold $W_1,\ldots,W_n$ and a calibration fold
$Z_1,\ldots,Z_B$, and pool the calibration scores with the test scores in place of the synthetic nulls.
Let $\widehat F_0$ be the empirical distribution function of the
estimation fold, and let $\widehat F_{\mathrm{all}}$ now denote that of
the $m+B$ pooled scores.
Of the $(m+B)\widehat F_{\mathrm{all}}(s)$ pooled scores not exceeding
$s$, about $B\widehat F_0(s)$ are calibration scores, so
$(m+B)\widehat F_{\mathrm{all}}(s)-B\widehat F_0(s)$ estimates the
number of test scores not exceeding $s$ without using the labels.
We take the smallest threshold at which this estimate reaches $K$,
\begin{equation}
\widehat s_{K,B,n}
=\inf\{s:(m+B)\widehat F_{\mathrm{all}}(s)-B\widehat F_0(s)\ge K\}.
\label{eq:split-selector}
\end{equation}
We then define the FDR level as in \eqref{eq:pooled-level}, with
$\widehat F_0(\widehat s_{K,B,n})$ in place of $\widehat t_{K,B}$:
\begin{equation}
\widehat q^{\mathrm{ref}}_K
=\chg{\left[\frac mK\,
\frac{1+B\widehat F_0(\widehat s_{K,B,n})}{B+1}\right]_{[q_{\min},q_{\max}]}}.
\label{eq:split-level}
\end{equation}
The level $\widehat q^{\mathrm{ref}}_K$ depends on the data only
through the estimation fold and the masked pool.

Once $\widehat q^{\mathrm{ref}}_K$ is obtained, we unmask the labels and
use the calibration fold to compute the conformal p-values and the
rejection set
\begin{equation}
\widetilde P_i=\frac{1+\sum_{b=1}^B\ind\{Z_b\le T_i\}}{B+1},
\quad
S^{\mathrm{ref}}=S_{\widehat q^{\mathrm{ref}}_K}(\widetilde P).
\label{eq:split-selection}
\end{equation}

The calibration size $B$ again balances finer conformal p-values
against a more variable estimated level, and the estimation fold adds a
further source of error.
Suppose that the p-values $P_i=F_0(T_i)$, which would be available if
$F_0$ were known, satisfy model \eqref{eq:model}.
In \cref{prop:variant-moments}, we show that, as $m\to\infty$ with $K$,
$B$ and $n$ of the same order as $m$,
\begin{equation}
\begin{aligned}
\E(R^{\mathrm{ref}})&=K+o(m^{1/2}),\\
\var(R^{\mathrm{ref}})&=
\frac{t_K(1-t_K)}{\{1-mt_Kf(t_K)/K\}^2}\,
(B+B^\star)^2\left(\frac1B+\frac1n\right)+o(m),
\end{aligned}
\label{eq:split-count-variance}
\end{equation}
where $R^{\mathrm{ref}}=|S^{\mathrm{ref}}|$ and $B^\star=mf(t_K)$.
For a fixed ratio $n/B$, the leading term of the variance is again
minimized at $B=B^\star$.
Since this value depends on unknown quantities \chg{such as $t_K$ and
$f(t_K)$}, we again use $B=m$ as a practical default.

\subsubsection{Conditional FDR control}

\begin{theorem}[Conditional FDR control]
\label{thm:split-reference}
Suppose that the test scores and the reference scores are independent,
and that the true null test scores and the reference scores have a
common continuous distribution function $F_0$.
Then, for any $B\ge1$ and $n\ge1$, the rejection set
$S^{\mathrm{ref}}$ and the level $\widehat q^{\mathrm{ref}}_K$ defined
by \eqref{eq:split-level} and \eqref{eq:split-selection} satisfy
\[
\E\{\FDP(S^{\mathrm{ref}})\mid\widehat q^{\mathrm{ref}}_K\}
\le\frac{m_0}{m}\,\widehat q^{\mathrm{ref}}_K
\quad\text{a.s}.
\]
\end{theorem}

Compared with \cref{thm:main-control}, the bound is unchanged although
the null distribution is estimated.
The proof is given in \cref{sec:proofs}.

{\printproofsfalse
\long\gdef\ProofSplitControl{%
\begin{proof}[of \cref{thm:split-reference}]
Let $\mathcal G^{\mathrm{ref}}$ be the sigma-field generated by the
estimation fold $W=(W_1,\ldots,W_n)$ and the masked pool, that is,
the unordered set of the $m+B$ test and calibration scores.
The level $\widehat q^{\mathrm{ref}}_K$ is
$\mathcal G^{\mathrm{ref}}$-measurable, so it suffices to show that, for
any $\mathcal G^{\mathrm{ref}}$-measurable level $\widehat q\in[0,1)$
and $S=S_{\widehat q}(\widetilde P)$,
$\E\{\FDP(S)\mid\mathcal G^{\mathrm{ref}}\}\le(m_0/m)\widehat q$.

Fix $i\in\mathcal H_0$, and let $X_{(1)}<\cdots<X_{(B+1)}$ denote the
ordered values of $T_i$ and $Z_1,\ldots,Z_B$, which are distinct almost
surely because $F_0$ is continuous.
Let $\rho_i$ be the rank of $T_i$ among them, and let $\mathcal F_i$ be
the sigma-field generated by $W$, $(T_j)_{j\ne i}$ and
$(X_{(1)},\ldots,X_{(B+1)})$.
Since $T_i$ and $Z_1,\ldots,Z_B$ are independent with distribution
function $F_0$, and independent of $W$ and $(T_j)_{j\ne i}$, the rank
$\rho_i$ is uniform on $\{1,\ldots,B+1\}$ and independent of
$\mathcal F_i$.
The estimation fold and the masked pool are functions of the variables
generating $\mathcal F_i$, so
$\mathcal G^{\mathrm{ref}}\subseteq\mathcal F_i$ and $\widehat q$ is
$\mathcal F_i$-measurable.

The conformal p-values in \eqref{eq:split-selection} depend on the
scores only through the indicators $\ind\{Z_b\le T_j\}$.
The remainder of the proof of \cref{thm:conditional-control} therefore
applies with $T_j$, $Z_b$ and $X_{(s)}$ in place of $P_j$, $U_b$ and
$W_{(s)}$, and gives
\[
\E\left\{\left.\frac{\ind\{i\in S\}}{R\vee1}\right|\mathcal F_i\right\}
\le\frac{\widehat q}{m}.
\]
Since $\mathcal G^{\mathrm{ref}}\subseteq\mathcal F_i$, the same bound
holds given $\mathcal G^{\mathrm{ref}}$, and summing over
$i\in\mathcal H_0$ gives the claim.

For the procedure with a null proportion estimate, let
$\mathcal H^{\mathrm{ref}}$ be the sigma-field generated by the
estimation fold, the non-null test scores and the unordered set of the
$m_0+B$ true null test scores and calibration scores.
Then
$\mathcal G^{\mathrm{ref}}\subseteq\mathcal H^{\mathrm{ref}}
\subseteq\mathcal F_i$, and, given $\mathcal H^{\mathrm{ref}}$, every
assignment of the values in that unordered set to the true null test
scores and the calibration scores is equally likely.
The proof of \cref{thm:adaptive-control} therefore applies with
$\mathcal G^{\mathrm{ref}}$ and $\mathcal H^{\mathrm{ref}}$ in place of
$\mathcal G$ and $\mathcal H$, and gives
$\E\{\FDP(S^{\mathrm{ada}})\mid\mathcal G^{\mathrm{ref}}\}
\le\widehat q$ for any $\mathcal G^{\mathrm{ref}}$-measurable level
$\widehat q\in[0,1)$.
\end{proof}
}
\ifprintproofs\ProofSplitControl\fi
}

\subsubsection{Asymptotic optimality}

The oracle level is $q_*=mt_K/K$, as in \cref{sec:bag-choice}, with
$t_K$ defined for the p-values $P_i=F_0(T_i)$.

\begin{proposition}[Asymptotic optimality]
\label{prop:split-targeting}
Suppose that the p-values $P_i=F_0(T_i)$ satisfy model \eqref{eq:model}
and \Cref{ass:regular}, and further assume that
$n/m\to\gamma\in(0,\infty)$.
Then, as $m\to\infty$,
$m^{1/2}(1-R^{\mathrm{ref}}/K)_+=O_p(1)$ and
\[
m^{1/2}\,\bigl\|\widehat q^{\mathrm{ref}}_K-q_*\bigr\|_2\to
\frac{\bigl(\kappa(1-\kappa)+t_\kappa(1-t_\kappa)
[\beta+\{\beta+f(t_\kappa)\}^2/\gamma]\bigr)^{1/2}}
{\kappa f(t_\kappa)}.
\]
If, in addition, $F$ is concave, then
\[
m^{1/2}\left\|\frac{R^{\mathrm{ref}}}K-1\right\|_2\to
\left(\frac1\beta+\frac1\gamma\right)^{1/2}
\frac{\{\beta+f(t_\kappa)\}\{t_\kappa(1-t_\kappa)\}^{1/2}}
{\kappa-t_\kappa f(t_\kappa)}.
\]
\end{proposition}

Compared with \cref{prop:targeting}, the oracle level is unchanged, and
both limits reduce to those of \cref{prop:targeting} as
$\gamma\to\infty$.

\subsubsection{Incorporating a null proportion estimate}

For a boundary $\ell$ that is a function of the estimation fold, with
$\widehat F_0(\ell)<1$, the analogue of \eqref{eq:pooled-pi0} is
\begin{equation}
\widehat\pi_0^{\mathrm{ref}}
=\min\left[1,\max\left\{0,
\frac{1+(m+B)\{1-\widehat F_{\mathrm{all}}(\ell)\}
-B\{1-\widehat F_0(\ell)\}}
{m\{1-\widehat F_0(\ell)\}}\right\}\right].
\label{eq:split-pi0}
\end{equation}
\chg{We multiply the raw level inside the brackets in \eqref{eq:split-level} by this estimate before clipping to $[q_{\min},q_{\max}]$, and apply the adaptive BH procedure \eqref{eq:adaptive-rule}}
at the resulting level to the conformal p-values in
\eqref{eq:split-selection}.
Under the assumptions of \cref{thm:split-reference}, the bound of
\cref{thm:adaptive-control} holds for the resulting rejection set and
level; the proof is given in \cref{sec:proofs}.

\section{Proofs}
\label{sec:proofs}

\subsection{Conditional FDR control}
\label{sec:sup-control}
\printproofstrue
\ProofOrdinaryControl
\chg{\Cref{thm:main-control} follows from \cref{thm:conditional-control}: $\widehat q_K$ depends on the data only through the masked pool, so it is $\mathcal G$-measurable, and the bound given $\mathcal G$ implies the bound given $\widehat q_K$.}

\begin{proof}[of \cref{rem:conservative-nulls}]
Let $\mathcal G$ be the sigma-field generated by the order statistics
of $(P_1,\ldots,P_m,U_1,\ldots,U_B)$.
The level $\widehat q_K$ is $\mathcal G$-measurable, so it suffices to
show that $\E\{\FDP(S)\mid\mathcal G\}\le(m_0/m)\widehat q$ for any
$\mathcal G$-measurable $\widehat q\in[0,1)$ and
$S=S_{\widehat q}(\widetilde P)$.

Fix $i\in\mathcal H_0$.
Let $W_{(1)}<\cdots<W_{(B+1)}$ be the order statistics of
$P_i,U_1,\ldots,U_B$, which are distinct almost surely, and let $\rho_i$
be the rank of $P_i$ among them.
Let $\mathcal F_i$ be the sigma-field generated by $(P_j)_{j\ne i}$ and
$(W_{(1)},\ldots,W_{(B+1)})$, so that
$\mathcal G\subseteq\mathcal F_i$.
Since $U_1,\ldots,U_B$ are independent and uniform on $(0,1)$ and
independent of $P_i$, and $(P_i,U_1,\ldots,U_B)$ is independent of
$(P_j)_{j\ne i}$,
\[
\Pr(\rho_i=r\mid\mathcal F_i)
=\frac{f_i(W_{(r)})}{\sum_{s=1}^{B+1}f_i(W_{(s)})}
\quad(r=1,\ldots,B+1),
\]
where the denominator is positive almost surely.
As $f_i$ is nondecreasing, these probabilities are nondecreasing in
$r$, so
\begin{equation}
\Pr(\rho_i\le k\mid\mathcal F_i)\le\frac k{B+1}
\quad(k=1,\ldots,B+1).
\label{eq:proof-rank-cons}
\end{equation}

Given $\mathcal F_i$, write $\widetilde P_j(r)$ and $R(r)$ for the
conformal p-values and the number of rejections when $\rho_i=r$.
Then $\widetilde P_i(r)=r/(B+1)$ and, for $j\ne i$,
\[
\sum_{b=1}^B\ind\{U_b\le P_j\}
=\sum_{s=1}^{B+1}\ind\{W_{(s)}\le P_j\}-\ind\{W_{(r)}\le P_j\},
\]
which is nondecreasing in $r$.
Hence every conformal p-value is nondecreasing in $r$, and $R(r)$ is
nonincreasing in $r$.
By the definition of the BH procedure, hypothesis $i$ is rejected when
$\rho_i=r$ if and only if
\begin{equation}
\frac r{B+1}\le\frac{\widehat q\,R(r)}m .
\label{eq:proof-reject-cons}
\end{equation}
The left side of \eqref{eq:proof-reject-cons} is increasing in $r$ and
the right side is nonincreasing, so \eqref{eq:proof-reject-cons} holds
exactly for $r\le r_*$, for some $r_*\in\{0,\ldots,B+1\}$.
If $r_*\ge1$,
\[
\begin{aligned}
\E\left\{\left.\frac{\ind\{i\in S\}}{R\vee1}\right|\mathcal F_i\right\}
&=\sum_{r=1}^{r_*}\frac{\Pr(\rho_i=r\mid\mathcal F_i)}{R(r)}
\le\frac{\Pr(\rho_i\le r_*\mid\mathcal F_i)}{R(r_*)}\\
&\le\frac{r_*}{(B+1)R(r_*)}\le\frac{\widehat q}{m},
\end{aligned}
\]
where the first inequality holds because $R(r)\ge R(r_*)$ for
$r\le r_*$, the second by \eqref{eq:proof-rank-cons}, and the third by
\eqref{eq:proof-reject-cons} at $r=r_*$.
If $r_*=0$ the left side is zero.
Since $\mathcal G\subseteq\mathcal F_i$, the same bound holds given
$\mathcal G$, and summing over $i\in\mathcal H_0$ gives
$\E\{\FDP(S)\mid\mathcal G\}\le(m_0/m)\widehat q$.
\end{proof}

\ProofAdaptiveControl

\ProofSplitControl

\subsection{Asymptotic optimality}
\label{sec:proof-targeting}
Assume model \eqref{eq:model} and \cref{ass:regular}.
Let $\widehat F_m(v)=m^{-1}\sum_i\ind(P_i\le v)$ and $\widehat F_B(v)=B^{-1}\sum_b\ind(U_b\le v)$.
Let $\kappa_m=K/m$, $b_m=B/m$, $g=f(t_K)$, $d_K=1-t_Kg/\kappa_m$, $h=t_K(1-t_K)$, $\eta=\widehat F_m(t_K)-\kappa_m$ and $\xi=\widehat F_B(t_K)-t_K$.
Then $\eta$ and $\xi$ are independent, with $\var(\eta)=\kappa_m(1-\kappa_m)/m$ and $\var(\xi)=h/B$.
Because $f$ is continuous at $t_\kappa$ with $f(t_\kappa)>0$, $F$ is strictly increasing near $t_\kappa$.
Because $K/m\to\kappa$, $t_K$ is therefore unique for large $m$ and $t_K\to t_\kappa$.
It follows that $g\to f(t_\kappa)$ and $d_K\to d=1-t_\kappa f(t_\kappa)/\kappa$, which is positive by \cref{ass:regular}.
We write $X_m=o_{L^2}(m^{-1/2})$ if $m^{1/2}\|X_m\|_2\to0$.

\begin{lemma}
\label{lem:prim-empirical}
Let $I$ be compact with $0<f_-\le f\le f_+$ on $I$, $v\in I$, $\delta_m=m^{-1/2}\log m$, $M_m\in\{\widehat F_m-F,\ \widehat F_B-\mathrm{id}\}$, and
\[
\begin{aligned}
\Delta_m&=\sup_u\max\{|\widehat F_m(u)-F(u)|,\,|\widehat F_B(u)-u|\},\\
G_m&=\sup\{w-u:(u,w]\subseteq I,\ \widehat F_m(w)=\widehat F_m(u)\},\\
\omega_m&=\sup_{|u-v|\le\delta_m}|M_m(u)-M_m(v)|.
\end{aligned}
\]
Then, for $x>0$ and $p\ge1$,
\[
\Pr(\Delta_m>x)\le C_0e^{-c_0mx^2},\quad
\|G_m\|_p=O\Bigl(\frac{\log m}m\Bigr),\quad
\|\omega_m\|_p=O(m^{-3/4}\log m).
\]
\end{lemma}

\begin{proof}[of \cref{lem:prim-empirical}]
By the Dvoretzky--Kiefer--Wolfowitz inequality and $B\ge\beta m/2$ for large $m$,
\[
\Pr(\Delta_m>x)\le2e^{-2mx^2}+2e^{-2Bx^2}\le4e^{-\min(1,\beta)mx^2},\quad
\E\Delta_m^p=\int_0^\infty px^{p-1}\Pr(\Delta_m>x)\,dx=O(m^{-p/2}).
\]
Divide $I$ into cells $J_1,\ldots,J_N$ of length $y/2$, with $N\le2|I|/y+1$.
Any subinterval of $I$ of length $y$ contains a cell, so
\[
\Pr(G_m>y)\le\sum_{k=1}^N\{1-F(J_k)\}^m\le\Bigl(\frac{2|I|}y+1\Bigr)e^{-mf_-y/2}.
\]
With $y_m=b\log m/m$ and $bf_-/2\ge p+1$, since $G_m\le|I|$,
\[
\|G_m\|_p\le y_m+|I|\Pr(G_m>y_m)^{1/p}\le y_m+Cm^{-1}=O(\log m/m).
\]
Let $n_m\in\{m,B\}$ be the sample size of $M_m$, and let $u_k=v+km^{-2}$ for $|k|\le m^2\delta_m$.
Then $n_m\{M_m(u_k)-M_m(v)\}$ is a sum of $n_m$ independent centred terms bounded by one, with variance $\sigma_m^2\le n_m(f_+\vee1)\delta_m$.
By Bernstein's inequality,
\[
\Pr\{|M_m(u_k)-M_m(v)|>x\}\le2\exp\Bigl\{-\frac{n_m^2x^2}{2(\sigma_m^2+n_mx/3)}\Bigr\}.
\]
By monotonicity of $\widehat F_m$ and $\widehat F_B$, for $u_k\le u\le u_{k+1}$,
\[
|M_m(u)-M_m(v)|\le\max_{l=k,k+1}|M_m(u_l)-M_m(v)|+(f_+\vee1)m^{-2}.
\]
Take $x_m=bm^{-3/4}\log m$.
Then $n_mx_m^2/\delta_m\asymp b^2\log m$ and $n_mx_m\gg\log m$, so a union bound over the at most $3m^2$ grid points gives
\[
\Pr(\omega_m>2x_m)\le12m^{2-cb^2}.
\]
Since $\omega_m\le2$, for $b$ large,
\[
\|\omega_m\|_p\le2x_m+2\Pr(\omega_m>2x_m)^{1/p}=O(m^{-3/4}\log m).
\]
\end{proof}

\begin{proof}[of \cref{prop:targeting}]
By continuity of $f$ at $t_\kappa$ and $f(t_\kappa)>0$, there is $\varepsilon>0$ such that \cref{lem:prim-empirical} applies on $I=[t_\kappa-3\varepsilon,t_\kappa+3\varepsilon]$.
For large $m$, $|t_K-t_\kappa|\le\varepsilon$ and $1-q_*f\ge d/2$ on $I$.
Let $E_m=\{\Delta_m\le x_0\}$, with $x_0$ small, so that $\Pr(E_m^c)\le Ce^{-cm}$.

By \eqref{eq:pooled-threshold}, $\widehat t_{K,B}$ is the first $v$ at which $\widehat F_m(v)+b_m\{\widehat F_B(v)-v\}$ reaches $\kappa_m$.
This function is within $C\Delta_m$ of $F$ uniformly and increases only through jumps of size $m^{-1}$, so
\begin{equation}
\widehat F_m(\widehat t_{K,B})+b_m\{\widehat F_B(\widehat t_{K,B})-\widehat t_{K,B}\}\in[\kappa_m,\kappa_m+m^{-1}).
\label{eq:prim-first}
\end{equation}
As $F$ is nondecreasing and strictly increasing on $I$, on $E_m$ we have
\begin{equation}
|\widehat t_{K,B}-t_K|\le C(\Delta_m+m^{-1}).
\label{eq:prim-cutoff}
\end{equation}
The raw level is $Q=(m/K)(1+B\widehat t_{K,B})/(B+1)$, so $|Q-q_*|\le C(\Delta_m+m^{-1})$ on $E_m$.
Since $q_{\min}<t_\kappa/\kappa<q_{\max}$, $\widehat q_K=Q$ on $E_m$.

The conformal p-value of $P_{(r)}=v$ is $\{1+B\widehat F_B(v)\}/(B+1)$, and $r=m\widehat F_m(v)$.
Hence $\widetilde P_{(r)}\le\widehat q_Kr/m$ if and only if
\[
D_m(P_{(r)})\le0,\quad
D_m(v)=\frac{1+B\widehat F_B(v)}{B+1}-\widehat q_K\widehat F_m(v),
\]
and $R$ is the largest such $r$.
On $E_m$, $\sup_v|D_m(v)-\{v-q_*F(v)\}|\le r_m=C(\Delta_m+m^{-1})$.
As $v-q_*F(v)$ vanishes at $t_K$ and has derivative $1-q_*f\ge d/2$ on $I$, $D_m<0$ on $(t_K-\varepsilon,t_K-2r_m/d)$.
Hence, on $E_m\cap\{G_m\le\varepsilon\}$, some $P_{(r)}\in[t_K-2r_m/d-G_m,\,t_K-2r_m/d)$ has $D_m(P_{(r)})\le0$, so $R\ge r$ and
\[
\kappa_m-\frac Rm\le F(t_K)-\widehat F_m(t_K-2r_m/d-G_m)\le f_+(2r_m/d+G_m)+\Delta_m.
\]
By \cref{lem:prim-empirical}, $\|(1-R/K)_+\|_2=O(m^{-1/2})$, which proves the first claim.

Let $E_m'=E_m\cap\{\Delta_m+G_m\le c\delta_m\}$, so that $\Pr(E_m'^c)\le Ce^{-c\log^2m}$.
Let $\varepsilon_m=\sup\{|f(u)-g|:|u-t_\kappa|\le2\delta_m+|t_K-t_\kappa|\}$, which tends to zero.
For a random argument $\widehat u$ with $|\widehat u-t_K|\le\delta_m$ on $E_m'$,
\begin{equation}
\widehat F_m(\widehat u)=\kappa_m+\eta+g(\widehat u-t_K)+e_m,\quad
\widehat F_B(\widehat u)=\widehat u+\xi+e_B,
\label{eq:prim-expansion}
\end{equation}
where, on $E_m'$, $|e_m|+|e_B|\le2\omega_m+\varepsilon_m|\widehat u-t_K|$, which is $o_{L^2}(m^{-1/2})$.
Substituting \eqref{eq:prim-expansion} with $\widehat u=\widehat t_{K,B}$ into \eqref{eq:prim-first} gives
\[
g(\widehat t_{K,B}-t_K)=-(\eta+b_m\xi)+o_{L^2}(m^{-1/2}).
\]
Hence
\[
Q-q_*=\frac{B}{\kappa_m(B+1)}(\widehat t_{K,B}-t_K)+O(m^{-1})
=-\frac{\eta+b_m\xi}{\kappa_mg}+o_{L^2}(m^{-1/2}).
\]
As $\widehat q_K$ and $Q$ are bounded, the events $E_m'^c$ and $\{\widehat q_K\ne Q\}$ contribute $o(m^{-1})$ to $\E(\widehat q_K-q_*)^2$.
Since $\var(\eta+b_m\xi)=\{\kappa_m(1-\kappa_m)+b_mh\}/m$,
\[
m\E(\widehat q_K-q_*)^2=\frac{\kappa_m(1-\kappa_m)+b_mh}{\kappa_m^2g^2}+o(1)
\to\frac{\kappa(1-\kappa)+\beta t_\kappa(1-t_\kappa)}{\kappa^2f(t_\kappa)^2}.
\]

If $F$ is concave, then $v-q_*F(v)$ is convex; as it vanishes at $t_K$ and increases on $I$, it is positive for $v>t_K$.
With the lower bound above, this gives $|s-t_K|\le2r_m/d+G_m$ on $E_m$, where $s=P_{(R)}$.
Let $s^+=P_{(R+1)}$.
Then $D_m(s)\le0<D_m(s^+)$, by maximality of $R$, and $\widehat F_m(s^+)=\widehat F_m(s)+m^{-1}$ give
\[
|D_m(s)|\le\widehat F_B(s^+)-\widehat F_B(s)\le(s^+-s)+2\omega_m\le G_m+2\omega_m=o_{L^2}(m^{-1/2}),
\]
because $s$ and $s^+\le s+G_m$ lie within $\delta_m$ of $t_K$ on $E_m'$.
Substituting \eqref{eq:prim-expansion} with $\widehat u=s$ and the expansion of $Q$ into $D_m(s)$ gives
\[
d_K(s-t_K)=\frac{t_K}{\kappa_m}\eta-\frac{\eta+b_m\xi}g-\xi+o_{L^2}(m^{-1/2}),
\]
where products of two first-order terms, each with $L^4$ norm $O(m^{-1/2})$, have $L^2$ norm $O(m^{-1})$ by the Cauchy--Schwarz inequality.
Since $R/m-\kappa_m=\eta+g(s-t_K)+e_m$ and $1+(gt_K/\kappa_m-1)/d_K=0$,
\[
\frac RK-1=-\frac{(b_m+g)\,\xi}{\kappa_md_K}+o_{L^2}(m^{-1/2}).
\]
Since $R\le m$, the event $E_m'^c$ contributes $o(1)$ to $m\E(R/K-1)^2$, so
\[
m\E\Bigl(\frac RK-1\Bigr)^2=\frac{(b_m+g)^2h}{b_m\kappa_m^2d_K^2}+o(1)
\to\frac{\{\beta+f(t_\kappa)\}^2t_\kappa(1-t_\kappa)}{\beta\{\kappa-t_\kappa f(t_\kappa)\}^2}.
\]
Taking square roots completes the proof.
\end{proof}

\subsection{Analysis of the number of rejections and the FDR level}
\label{sec:sup-variant-moments}

We treat together the procedure of \cref{sec:method}, the adaptive
procedure of \cref{sec:adaptive-extension} and their conformal
counterparts in \cref{sec:conformal-extension}.
In the last case we take $P_i=F_0(T_i)$ and $U_b=F_0(Z_b)$.

Let $P_1,\ldots,P_m$ be independent with distribution functions
$F_1,\ldots,F_m$, and let $F=m^{-1}\sum_{i=1}^mF_i$ have density $f$.
For the adaptive procedure let $\rho$, $w$ and $c_0$, $c_1$, $c_2$ be
as in \eqref{eq:adaptive-coefficients}; \chg{for} the non-adaptive procedure, let $\rho=1$ and
$w=0$, so that $c_j=f(t_K)^jt_K(1-t_K)$ for $j=0,1,2$.
For $v\in(0,1)$ let
\[
\eta(v)=\sum_{i=1}^m\ind\{P_i\le v\}-mF(v),\quad
\xi(v)=\sum_{b=1}^B\ind\{U_b\le v\}-Bv,
\]
and, with a null reference sample, let
$\zeta(v)=n^{-1}\sum_{j=1}^n\ind\{F_0(W_j)\le v\}-v$; otherwise let
$\zeta(v)=0$.
Then $\eta$, $\xi$ and $\zeta$ are independent with mean zero and, for
$u\le v$,
\begin{equation}
\operatorname{cov}\{\xi(u),\xi(v)\}=Bu(1-v),\quad
\operatorname{cov}\{\zeta(u),\zeta(v)\}=u(1-v)/n.
\label{eq:reference-covariance}
\end{equation}

\begin{assumption}
\label{ass:expansion}
As $m\to\infty$,
\begin{enumerate}
\item[(a)] $K/m$, $B/m$ and, with a null reference sample, $n/m$ are
bounded away from zero and infinity, $K/m$ is bounded away from one,
and $t_K$, $f(t_K)$ and $1-B^\star t_K/K$ are bounded away from zero;
\item[(b)] $F$ has a bounded second derivative near $t_K$ and, for the
adaptive procedure, near $\lambda$, with $\lambda-t_K$, $\rho$ and
$1-\rho$ bounded away from zero and $\tau=\lambda+O(B^{-1})$;
\item[(c)] for $\widehat v\in\{\widehat t,s\}$, where $\widehat t$ is
the threshold selected for the level and $s$ the largest rejected
p-value,
\begin{gather*}
\E\{(\widehat v-t_K)^4\}=O(m^{-2}),\\
\|\eta(\widehat v)-\eta(t_K)\|_2+\|\xi(\widehat v)-\xi(t_K)\|_2
+B\|\zeta(\widehat v)-\zeta(t_K)\|_2=o(m^{1/2});
\end{gather*}
and, for the adaptive procedure,
\[
\E\{(\widehat\pi_0^{\mathrm{pool}}-\rho)^4\}
+\E\{(\widehat\pi_0^{\mathrm{conf}}-\rho)^4\}=O(m^{-2}),
\]
and the second display holds with $t_K$ replaced by $\lambda$ and
$\widehat v$ by each threshold at which
$\widehat\pi_0^{\mathrm{pool}}$ and $\widehat\pi_0^{\mathrm{conf}}$
are evaluated.
\end{enumerate}
\end{assumption}

\chg{For the procedure of \cref{sec:method}, part (c) holds under model \eqref{eq:model} and \cref{ass:regular} with $F$ concave, by \cref{lem:prim-empirical} and the argument in the proof of \cref{prop:targeting}.}

\begin{proposition}[Expansions for the procedure and its extensions]
\label{prop:variant-moments}
Let $\widehat q$ be the level before capping and $R$ the number of rejections at
that level.
Under \cref{ass:expansion}, $\E(R)=K+o(m^{1/2})$,
$\E(\widehat q)=\rho q_*+o(m^{-1/2})$,
\begin{align}
\var(R)&=\frac{1+B/n}{(1-B^\star t_K/K)^2}
\left(c_0B+2mc_1+\frac{m^2c_2}{B}\right)+o(m),
\label{eq:unified-count-variance}\\
\var(\widehat q)&=\frac{\rho^2}{\{Kf(t_K)\}^2}
\biggl[\var\{\eta(t_K)+w\eta(\lambda)\}+c_0B+\frac{c_0B^2+2mc_1B+m^2c_2}{n}\biggr]+o(m^{-1}),
\label{eq:unified-level-variance}
\end{align}
where
\begin{align*}
\var\{\eta(t_K)+w\eta(\lambda)\}
&=\sum_{i=1}^m\bigl[F_i(t_K)\{1-F_i(t_K)\}
+2wF_i(t_K)\{1-F_i(\lambda)\}+w^2F_i(\lambda)\{1-F_i(\lambda)\}\bigr].
\end{align*}
\end{proposition}

\begin{proof}
Throughout the proof, remainder terms in expansions of random
quantities are in $L^2$ norm, and
$\widehat\pi_0^{\mathrm{pool}}$ and $\widehat\pi_0^{\mathrm{conf}}$ are
replaced by one for the non-adaptive procedure.
By \cref{ass:expansion}(c), the relation
\[
\sum_{i=1}^m\ind\{P_i\le\widehat t\}
+\sum_{b=1}^B\ind\{U_b\le\widehat t\}
-B\{\widehat t+\zeta(\widehat t)\}=K+O(1)
\]
defining $\widehat t$ gives
\[
B^\star(\widehat t-t_K)+\eta(t_K)+\xi(t_K)-B\zeta(t_K)=o(m^{1/2}),
\]
and, since $\zeta(\widehat t)=\zeta(t_K)+o(m^{-1/2})$,
\[
\frac{\widehat q}{\widehat\pi_0^{\mathrm{pool}}}-q_*
=\frac{-\eta(t_K)-\xi(t_K)+(B^\star+B)\zeta(t_K)}{Kf(t_K)}
+o(m^{-1/2}).
\]
For the adaptive procedure, the same expansions at $\lambda$, together
with the fact that $\widetilde P_i\ge\tau$ if and only if $P_i$ is at
least the $(k_\tau-1)$th smallest of $U_1,\ldots,U_B$, give
\begin{align*}
\widehat\pi_0^{\mathrm{pool}}-\rho
&=\frac{-\eta(\lambda)-\xi(\lambda)+\{mf(\lambda)+B\}\zeta(\lambda)}
{m(1-\lambda)}+o(m^{-1/2}),\\
\widehat\pi_0^{\mathrm{conf}}-\rho
&=\frac{-\eta(\lambda)+\{mf(\lambda)/B\}\xi(\lambda)}{m(1-\lambda)}
+o(m^{-1/2}),
\end{align*}
so that
\begin{equation}
\widehat\pi_0^{\mathrm{pool}}-\widehat\pi_0^{\mathrm{conf}}
=\frac{1+mf(\lambda)/B}{m(1-\lambda)}
\{B\zeta(\lambda)-\xi(\lambda)\}+o(m^{-1/2}).
\label{eq:tail-noise-cancellation}
\end{equation}
By the fourth-moment bounds in \cref{ass:expansion}(c), the product of
$\widehat q/\widehat\pi_0^{\mathrm{pool}}-q_*$ and
$\widehat\pi_0^{\mathrm{pool}}-\rho$ is $O(m^{-1})$, so
\begin{equation}
\begin{aligned}
\frac{Kf(t_K)}{\rho}(\widehat q-\rho q_*)
={}&-\eta(t_K)-w\eta(\lambda)-\xi(t_K)-w\xi(\lambda)
+(B^\star+B)\zeta(t_K)\\
&+w\{mf(\lambda)+B\}\zeta(\lambda)+o(m^{1/2}).
\end{aligned}
\label{eq:unified-level-expansion}
\end{equation}

By \cref{ass:expansion}(c), the threshold $s$ satisfies
\[
s+\frac{\xi(t_K)}{B}
=\frac{\widehat q\{K+\eta(t_K)+B^\star(s-t_K)\}}
{m\,\widehat\pi_0^{\mathrm{conf}}}+o(m^{-1/2}).
\]
Linearizing about $\widehat q=\rho q_*$ and $\widehat\pi_0^{\mathrm{conf}}=\rho$,
\begin{align*}
\left(1-\frac{B^\star t_K}{K}\right)(s-t_K)
={}&\frac{K(\widehat q-\rho q_*)}{m\rho}
-\frac{t_K(\widehat\pi_0^{\mathrm{conf}}-\rho)}{\rho} +\frac{t_K\eta(t_K)}{K}-\frac{\xi(t_K)}{B}+o(m^{-1/2}).
\end{align*}
As $R-K=\eta(t_K)+B^\star(s-t_K)+o(m^{1/2})$, substituting
\eqref{eq:unified-level-expansion} and
\eqref{eq:tail-noise-cancellation} gives
\begin{equation}
\begin{aligned}
\left(1-\frac{B^\star t_K}{K}\right)(R-K)
={}&\left(1+\frac{B^\star}{B}\right)\{B\zeta(t_K)-\xi(t_K)\}+w\left\{1+\frac{mf(\lambda)}{B}\right\}
\{B\zeta(\lambda)-\xi(\lambda)\}+o(m^{1/2}).
\end{aligned}
\label{eq:unified-count-expansion}
\end{equation}

By \eqref{eq:reference-covariance}, $B\zeta(u)-\xi(u)$ and
$B\zeta(v)-\xi(v)$ have covariance $(B+B^2/n)u(1-v)$ for $u\le v$, so
the leading term on the right of \eqref{eq:unified-count-expansion} has
variance
\[
\left(B+\frac{B^2}{n}\right)
\left(c_0+\frac{2mc_1}{B}+\frac{m^2c_2}{B^2}\right),
\]
which gives the leading term of \eqref{eq:unified-count-variance}.
The terms of \eqref{eq:unified-level-expansion} in $\eta$, $\xi$ and
$\zeta$ are independent, with variances
$\var\{\eta(t_K)+w\eta(\lambda)\}$, $c_0B$ and
$(c_0B^2+2mc_1B+m^2c_2)/n$, which gives the leading term of
\eqref{eq:unified-level-variance}.
The leading terms have mean zero, and variances of order $m$ and
$m^{-1}$ respectively, so the stated means, variances and mean squared
errors follow from the Cauchy--Schwarz inequality.
\end{proof}

\section{Additional numerical results}
\label{sec:simulation-details}
\label{sec:validation-study}


\subsection{Comparison methods}
\label{sec:comparators}

\begin{table}[!htb]
\caption{Guarantees of methods targeting $K$ discoveries.}
\label{tab:comparison}
\centering
\begin{tabular}{lll}
\toprule
Method & Guarantee at the reported level & Rejections \\
\midrule
BH inversion & None & At least $K$ \\
\chg{FDP bounds (converted)} & \chg{Marginal: $\E\{\FDP/\widehat q_K\}\le1$} & \chg{Data-dependent} \\
E-values-based & Marginal: $\E\{\FDP/\widehat q_K\} \le 1$ & \chg{Data-dependent} \\
Proposal & Conditional:
  $\E\{\FDP \mid \widehat q_K\} \le \widehat q_K$ & Close to $K$ \\
\bottomrule
\end{tabular}

\smallskip
{\footnotesize \chg{The lower bound for BH inversion assumes $K$ is attainable within the acceptable range; the e-value and FDP-bound methods select levels from ranks within 10\% of $K$, but their final counts depend on the data. `Close to $K$\textquotesingle{} is made precise in \cref{prop:targeting}.}}
\end{table}

\subsubsection{Oracles}

\chg{In the numerical comparisons, all reported levels are clipped to $[q_{\min},q_{\max}]$.}
The population oracle applies the BH procedure at the level $mt_K/K$,
where $F(t_K)=K/m$.
The independent-sample oracle applies it at the level
$mP^{\mathrm{pilot}}_{(K)}/K$, where $P^{\mathrm{pilot}}_{(K)}$ is the
$K$th smallest p-value in an independent copy of the data.
The adaptive population oracle reports the level $\pi_0mt_K/K$ and
applies the BH procedure at that level divided by the known null
proportion $\pi_0$.

\subsubsection{\chg{BH inversion}}

BH inversion reports the level $\min_{r\ge K}mP_{(r)}/r$, computed from
the p-values that are then tested; its adaptive version multiplies
this level by the estimate of the null proportion of
\citet{StoreyTaylorSiegmund2004} with a fixed $\lambda$.

\phantomsection\label{sec:bh-inversion-null}%
BH inversion does not control FDR at the reported level. Below is a simple counter example.
Suppose that all $m$ hypotheses are null, with independent uniform p-values, and consider BH inversion with $K=1$ and no cap, which reports $\widehat q=\min_{1\le r\le m}mP_{(r)}/r$ and applies the BH procedure at $\widehat q$.
If $r^*$ attains the minimum, then $P_{(r^*)}=\widehat q\,r^*/m$, so the BH procedure rejects at least $r^*\ge1$ hypotheses, all of them null, and $\FDP(S)=1$.
Moreover, $\widehat q\le q$ exactly when the BH procedure at level $q$ makes a rejection, which under the global null has probability equal to its FDR, $q$ \citep{BenjaminiHochberg1995}.
Hence $\widehat q$ is uniform on $(0,1)$ and
\[
\E\{\FDP(S)\mid\widehat q\}=1>\widehat q\quad\text{a.s.}
\]
With a cap $q_{\max}$ and any $K$, the same holds whenever the reported level is below $q_{\max}$.

\subsubsection{E-value-based methods}
\label{sec:e-comparator}

For a positive function $h$, statistics $X_1,\ldots,X_m$ and
independent null statistics $Z_1,\ldots,Z_B$, let
\[
E_i=\frac{(B+1)h(X_i)}{h(X_i)+\sum_{b=1}^Bh(Z_b)}
\quad(i=1,\ldots,m).
\]
For $i\in\mathcal H_0$, $X_i,Z_1,\ldots,Z_B$ are exchangeable, so
$\E(E_i)=1$.
Let $E_{(1)}\ge\cdots\ge E_{(m)}$.
The candidate level is the minimum of $m/(rE_{(r)})$ over ranks $r$
with $|r-K|\le0.1K$ at the ends of tie blocks, with ties broken by the
distance to $K$ and then by the smaller rank.
\chg{The reported level $\widehat q$ is the candidate level clipped to $[q_{\min},q_{\max}]$, and the e-BH procedure \citep{WangRamdas2022} is applied at $\widehat q$.}
For any rejection set $S$ and level $\widehat q$ such that
$E_i\ge m/(\widehat q\,|S|)$ for all $i\in S$,
\begin{equation}
\frac{\FDP(S)}{\widehat q}\le\frac1m\sum_{i\in\mathcal H_0}E_i,
\quad
\E\left\{\frac{\FDP(S)}{\widehat q}\right\}\le\frac{m_0}{m},
\label{eq:e-envelope}
\end{equation}
where the ratio is zero if $S=\emptyset$.

The first e-value-based method uses the fixed function
\[
h(x)=\left\{\frac{e^{-x-1/2}+e^{-3x-9/2}}2\right\}^{\gamma},
\]
with $\gamma=3$ in the simulations and $\gamma=50$ for the
breast-cancer data, following the powered-odds construction of
\citet{ZhuSimeone2026}.
Here $X_i$ is the normal score in the simulations and
$\Phi^{-1}(P_i)$ for the breast-cancer data.

\phantomsection\label{sec:source-odds}%
The second method learns $h$ from the data.
The statistics $X_1,\ldots,X_m$ are divided into five folds
independently of their values, and $m$ further null statistics,
independent of $Z_1,\ldots,Z_B$, are generated for training.
\chg{For each fold, a linear discriminant with equal prior probabilities, pooled variance and slope constrained to be nonpositive is fitted to the statistics outside the fold against the training nulls.}
The fitted odds raised to the power $\gamma$ are used as $h$ for the
statistics in the fold and for $Z_1,\ldots,Z_B$.

\subsubsection{FDP-bound-based methods}

Let $S_r=\{i:P_i\le P_{(r)}\}$ for ranks $r$ at the ends of tie blocks,
so that $|S_r|=r$, and, for $\delta\in(0,1)$, let
\[
c_\delta=\frac{\log(1/\delta)}{\log\{1+\log(1/\delta)\}},
\quad
U_\delta(S_r)
=\min\left\{1,\frac{\lfloor c_\delta(1+mP_{(r)})\rfloor}{r}\right\}.
\]

\begin{lemma}[\citealp{KatsevichRamdas2020}, Theorem~1]
\label{lem:sorted-envelope}
Suppose that the null p-values are independent and stochastically no
smaller than uniform.
If $0<\delta\le0.3$, then
\[
\Pr\{\FDP(S_r)\le U_\delta(S_r)\text{ for all }r\}\ge1-\delta.
\]
\end{lemma}

\begin{proposition}
\label{prop:envelope-conversion}
Let $\mathcal S$ be a family of sets and $\delta\in(0,1)$, and suppose
that $U_\delta(S)\in[0,1]$ satisfies
$\Pr\{\FDP(S)\le U_\delta(S)\text{ for all }S\in\mathcal S\}
\ge1-\delta$.
Let $S\in\mathcal S\cup\{\emptyset\}$ and $\widehat q$ depend on the
data, with $\widehat q\ge U_\delta(S)+\delta^{1/2}$ whenever
$S\ne\emptyset$.
Then
\[
\E\left\{\frac{\FDP(S)}{\widehat q}\right\}\le1,
\]
where the ratio is zero if $S=\emptyset$.
\end{proposition}

\begin{proof}
Let $A$ be the event that $\FDP(S')\le U_\delta(S')$ for all
$S'\in\mathcal S$.
If $S\ne\emptyset$, then on $A$
\[
\frac{\FDP(S)}{\widehat q}
\le\frac{U_\delta(S)}{U_\delta(S)+\delta^{1/2}}
\le\frac1{1+\delta^{1/2}},
\]
as $U_\delta(S)\le1$, and on $A^{\rm c}$ the ratio is at most
$1/\widehat q\le\delta^{-1/2}$.
Since $\Pr(A^{\rm c})\le\delta$,
\[
\E\left\{\frac{\FDP(S)}{\widehat q}\right\}
\le\frac{1-\delta}{1+\delta^{1/2}}+\frac{\delta}{\delta^{1/2}}=1.
\]
\end{proof}

The converted bound controls the expected ratio of the FDP to the reported level, but it does not provide the conditional
guarantee \eqref{eq:target-guarantee}.

In the comparisons, for each bound we minimize $U_\delta(S_r)+\delta^{1/2}$ over ranks $r$ at the ends of tie blocks with $|r-K|\le0.1K$, \chg{clip it to $[q_{\min},q_{\max}]$ to give the reported level $\widehat q$}, and reject $S_r$ for the rank $r$ closest to $K$ with $U_\delta(S_r)+\delta^{1/2}\le\widehat q$; if there is no such rank, no rejection is made.

\begin{figure}[!htb]
\centering
\includegraphics[width=\linewidth]{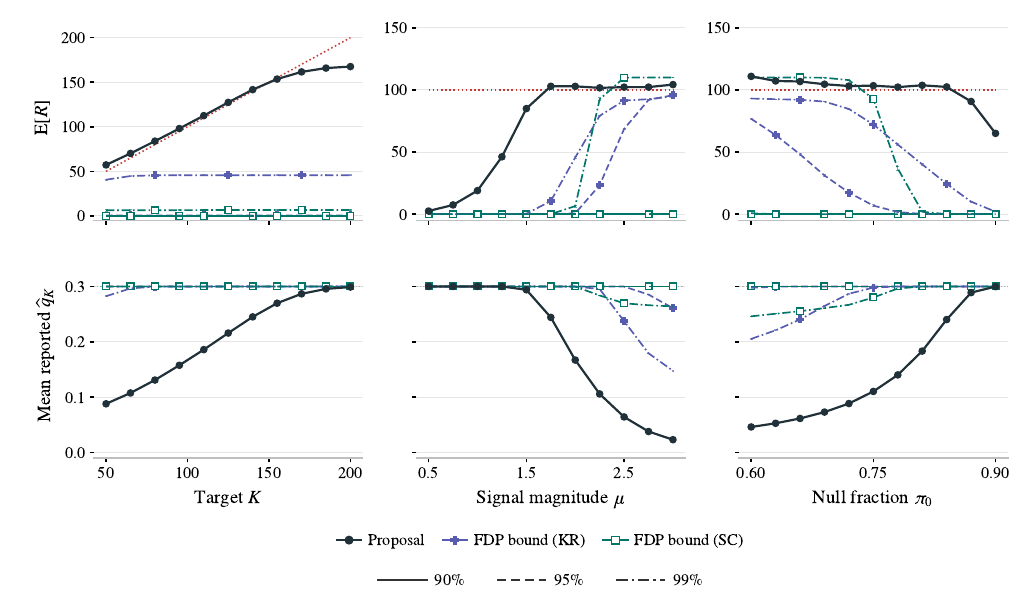}
\caption{Converted FDP bounds at 90\% (solid), 95\% (dashed) and 99\% (dash-dot) confidence, \chg{in the first three sweeps of \cref{fig:simulation-panels}, with levels clipped to $[0.01,0.30]$}: mean number of rejections (top) and mean reported level (bottom).
Conversion adds $\delta^{1/2}$ to the bound (\cref{prop:envelope-conversion}); \chg{at 90\% this exceeds the cap, so the bound-based methods make no rejections.}
Red lines mark the target $K$; bars show $1.96$ Monte Carlo standard errors.}
\label{fig:fdp95-sweeps}
\end{figure}

\begin{figure}[!htb]
\centering
\includegraphics[width=\linewidth]{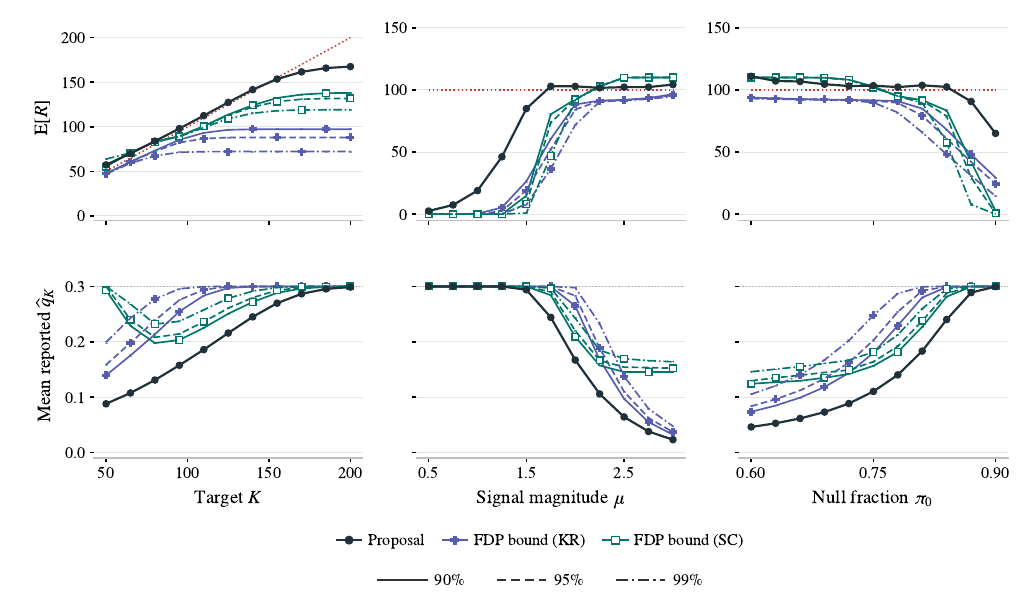}
\caption{As \cref{fig:fdp95-sweeps}, without conversion; these bounds give simultaneous confidence statements for the FDP, not FDR control at the reported level.}
\label{fig:fdp95-native}
\end{figure}

\phantomsection\label{sec:fdp-sensitivity}%
The main comparisons use 99\% confidence, chosen after seeing the results as the level among 90, 95 and 99\% with the smallest mean $|R-K|$ at the default setting (this choice favors the FDP-bound-based methods).
\Cref{fig:fdp95-sweeps,fig:fdp95-native} compare the three confidence levels with and without conversion.
\chg{The retained main and benchmark KR simulations use the original floating-point comparison at the cap, whereas the sensitivity calculation includes equality at the cap; their 99\% mean rejection counts at the default setting are therefore 43.9 and 45.6, respectively.}

\subsection{Choice of synthetic null size $B$}
\label{sec:variance-study}
\label{sec:extension-study}

We extend the study of the number $B$ of synthetic nulls in \cref{sec:method} to the adaptive and conformal extensions, whose variances are approximated in \cref{prop:variant-moments}.
We use the default setting of \cref{sec:numerical} with $B/m\in\{0.25,0.5,1,2,5\}$, $\lambda=0.5$, $\tau=\lfloor(B+1)/2\rfloor/(B+1)$, $n=B$ \chg{and uncapped levels, with 3000 independent repetitions; within each repetition, the original data and nested prefixes of the calibration and estimation samples are shared across values of $B$.}

\begin{figure}[!htb]
\centering
\includegraphics[width=.30\linewidth]{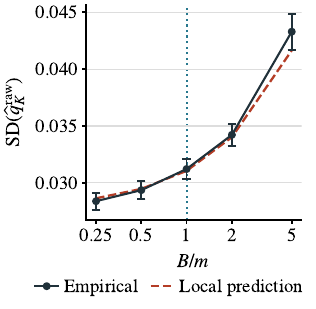}
\caption{Standard deviation of the raw level of the proposal in the setting of \cref{sec:variance-study}.
Bars are approximate 95\% Monte Carlo intervals, the dashed curve is the theoretical approximation, and the dotted line marks the recommended $B=m$.}
\label{fig:b-raw-q-variance}
\end{figure}

\begin{figure}[!htb]
\centering
\includegraphics[width=.60\linewidth]{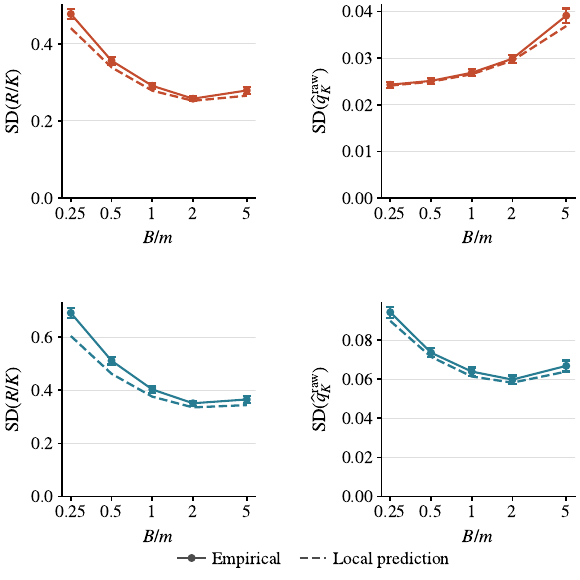}
\caption{\chg{Standard deviations of $R/K$ (left) and the raw level (right) for the adaptive (top row) and conformal (bottom row) extensions, with the design and interval convention of \cref{fig:b-raw-q-variance}.}}
\label{fig:validation-variants}
\end{figure}

As shown in \cref{fig:validation-variants}, the observations are similar to those for the proposal in \cref{fig:b-raw-q-variance}: the theoretical approximations track the empirical standard deviations across $B$.

\subsection{Simulations for the extensions}
\label{sec:additional-comparisons}
We extend the comparisons of \cref{sec:numerical} to the adaptive and conformal extensions in \cref{sec:adaptive-extension,sec:conformal-extension}.
The comparison methods are the same as in \cref{sec:numerical}, and the design is similar, with $\lambda=0.5$ for the adaptive extension and $n=B$ for the conformal extension; see the caption of \cref{fig:validation-all-methods}.
\Cref{fig:validation-levels} shows the comparison methods.

\subsubsection{Adaptive}

As shown in \cref{fig:validation-all-methods}, the results are largely similar to those in \cref{sec:numerical}; the difference is that the adaptive extension reports smaller levels, by about the factor $\rho$ of \cref{prop:adaptive-targeting}.

\subsubsection{Conformal}

\begin{figure}[!htb]
\centering
\includegraphics[width=\linewidth]{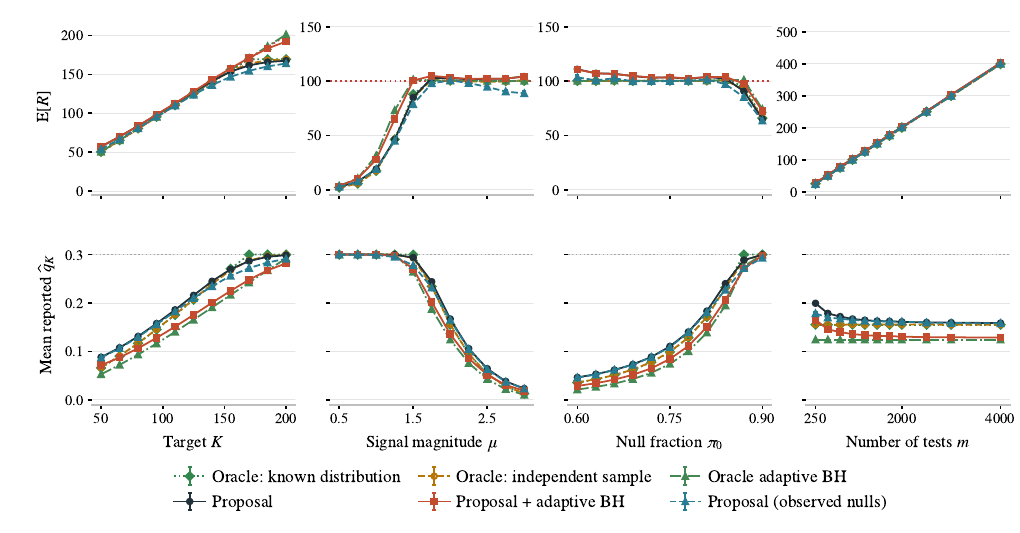}
\caption{Mean number of rejections (top) and mean reported level (bottom) of the proposal, its adaptive and conformal extensions, and the oracles.
Each column varies one parameter from the default of \cref{sec:numerical}: $K=50,65,\ldots,200$, $\mu=0.5,0.75,\ldots,3$, $\pi_0=0.60,0.63,\ldots,0.90$, \chg{or $m\in\{250,500,750,1000,1250,1500,1750,2000,2500,3000,4000\}$ with $K=m/10$ and $B=m$; levels are clipped to $[0.01,0.30]$.}
Red lines mark the target $K$; bars show $1.96$ Monte Carlo standard errors.}
\label{fig:validation-all-methods}
\end{figure}

\begin{figure}[!htb]
\centering
\includegraphics[width=\linewidth]{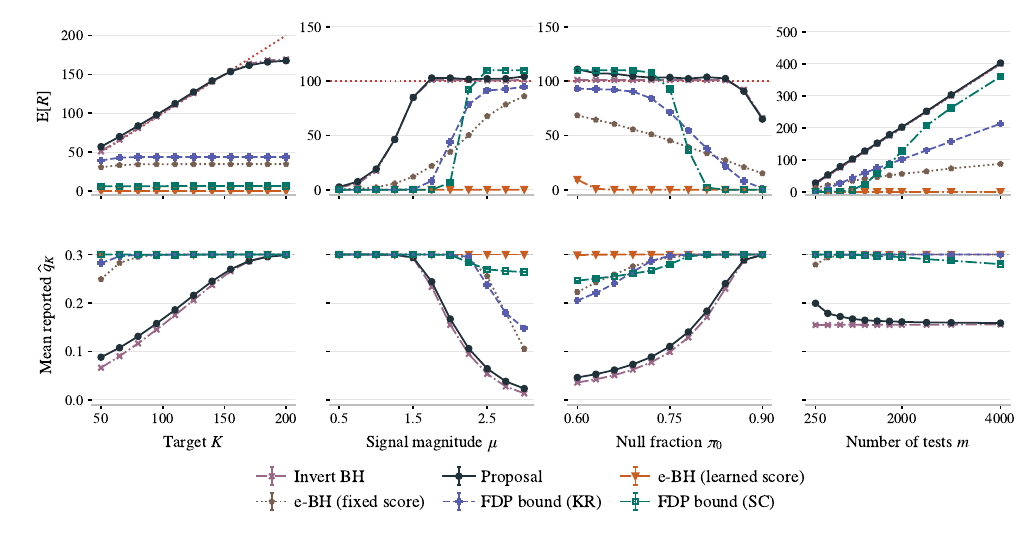}
\caption{As \cref{fig:validation-all-methods}, for the comparison methods of \cref{sec:comparators}; \chg{KR and SC use converted 99\% bounds for original p-values, and BH inversion has no FDR guarantee.}}
\label{fig:validation-levels}
\end{figure}

\chg{In \cref{fig:validation-all-methods}, the conformal extension (labelled Proposal (observed nulls)) averages 100.6 discoveries at mean level 0.165 at the default setting, compared with 102.9 and 0.167 for the proposal.}

\subsection{Mouse-protein data}
\label{sec:protein-application}
We analyse a mouse-brain proteomics dataset with 1040 candidate proteins and 421 annotated reference proteins, used as nulls.
The score of each protein is the negative of its second treatment-to-control abundance ratio in the prefrontal cortex, so small scores indicate enrichment.
The reference proteins are split at random into an estimation and a calibration set; see the caption of \cref{tab:mouse-pfc2}.
We apply the conformal extension of \cref{sec:conformal-extension}; the e-value methods use $\gamma=50$, with the learned score trained on the 121 estimation references, and the bound of \citet{SongJinCandes2026} is calibrated for the dependent conformal p-values.

\Cref{tab:mouse-pfc2} gives the results.
\chg{The proposal makes 59 discoveries at reported level 0.20, close to the target of 60; BH inversion makes 63 at level 0.22, without an FDR guarantee;} e-BH with the fixed and learned scores makes 13 and 0 discoveries, and the FDP bound makes none, all at the cap 0.30.

\begin{table}[!htb]\centering\tablefont
\caption{PFC2 mouse proteins at \chg{$K=60$} using one reference partition with seed 318: 1040 tests, 121 estimation and 300 calibration references.
Reported levels are clipped to $[0.01,0.30]$.}
\label{tab:mouse-pfc2}
\begin{tabular}{@{}l*{5}{c}@{}}\toprule
Metric & \chg{BH inversion} & Proposal & \begin{tabular}[c]{@{}c@{}}e-BH\\(learned score)\end{tabular} & \begin{tabular}[c]{@{}c@{}}e-BH\\(fixed score)\end{tabular} & \begin{tabular}[c]{@{}c@{}}FDP bound\\(SC)\end{tabular}\\\midrule
\chg{Reported $\widehat q_{60}$} & \chg{0.22} & \chg{0.20} & 0.30 & 0.30 & 0.30 \\
Rejections & \chg{63} & \chg{59} & 0 & 13 & 0 \\
\bottomrule\end{tabular}
\par\smallskip\begin{minipage}{0.96\linewidth}\footnotesize
\end{minipage}
\end{table}



\end{document}